\documentclass[twocolumn,showpacs]{revtex4}

\usepackage{epsfig}

\usepackage{amsmath}
\usepackage{amsfonts}
\usepackage{amssymb}

\begin{document}

\title{The role of clustering and gridlike ordering in epidemic spreading}
\author{Thomas Petermann$^1$}
\email{Thomas.Petermann@epfl.ch}
\author{Paolo De Los Rios$^{1}$}
\address{$^1$Laboratoire de Biophysique Statistique, ITP - FSB, Ecole Polytechnique F\'ed\'erale de Lausanne, CH-1015, Lausanne, Switzerland.}
\date{\today}

\begin{abstract}

The spreading of an epidemic is determined by the connectiviy patterns which underlie the population. While it has been noted that a virus spreads more easily on a network in which global distances are small, it remains a great challenge to find approaches that unravel the precise role of local interconnectedness. Such topological properties enter very naturally in the framework of our two-timestep description, also providing a novel approach to tract a probabilistic system. The method is elaborated for SIS-type epidemic processes, leading to a quantitative interpretation of the role of loops up to length 4 in the onset of an epidemic.

\end{abstract}
\pacs{89.75-k Complex systems - 05.10.-a Computational methods in statistical physics and nonlinear dynamics - 87.23.-n Ecology and evolution}
\maketitle

%\narrowtext
\section{Introduction}

Almost all of us have already met someone far from home, who turns out to share a friend with us. Milgram first put this social phenomenon on a firm basis, finding that any two individuals are separated, on average, by six acquaintances   \cite{milgram}. Furthermore, our society is composed of groups within which, if individual A is acquainted with persons B and C, then B is likely to know C. In the language of social networks, this means that the individuals A, B and C are arranged on a triangle. These two facts tell us that in the social universe, global distances are small, and local interconnectedness, i.e.\ {\it clustering}, is high. A network that possesses both of these topological properties is called a {\it small world}. In order to find a model that accounts for both of these properties, one could choose a regular lattice whose nodes are indeed locally highly interconnected. But since global distances are large in this type of network, it is an ineligible candidate for a small world model. In a random graph \cite{erdoes}, that is a set of nodes (e.g.\ laid out on a virtual circle) with connections between them established at random, a link is more likely to point to a far away node than to one close by. (Here we refer to the distance on the virtual circle's circumference.) As a consequence, the presence of the very many long-range links account for the desired global property, but the degree of clustering is low. Watts and Strogatz combined these two insights and proposed a model that interpolates between a regular lattice and a random graph, thus capturing both the global and local topological properties mentioned above \cite{watts}. But the small world property is not merely exhibited by social networks. Yet, through the increased availability of data, it was found that the simultaneous occurence of high local {\it and} global interconnectedness is prominent to a much wider class of systems, notably  the Internet \cite{pastor2,yook}, the World Wide Web \cite{albert2}, metabolic networks \cite{wagner} and food webs \cite{montoya}. 

In addition to the small world property, there is another important fact when it comes to characterizing the topology of a complex network: not all the nodes have the same number of edges. The corresponding measure is the {\it degree distribution} $P(k)$ which gives the probability that a randomly chosen node has degree $k$, that is $k$ edges. For most of the just mentioned examples, this distribution was found to be $P(k) \sim k^{-\gamma}$, $2 < \gamma \leq 3$, implying the absence of a characteristic (degree) scale, hence the name {\it scale-free} network. This emergence of scaling can be understood, for example,  in terms of {\it growing} networks. Starting from a small core graph, at each time step a node is added together with a certain number of edges that are connected to existing nodes, the latter being chosen {\it preferentially}, that is a link is more likely established to a high degree node \cite{barabasi}. This fat-tailed degree distribution implies the presence of {\it hubs} (high degree nodes) which hold together the network and play a crucial role in issues such as robustness or fragility \cite{albert,dorogovtsev}.

This explosion of research activity in the field of complex networks has also shed new light on epidemic spreading since the latter can be regarded as a dynamical process occuring on a complex network: a computer virus spreads on the Internet, and HIV (as a biological example) propagates on top of the web of human sexual contacts. Also these two examples fall into the category of scale-free networks \cite{faloutsos,liljeros}, and it is this topological property that accounts for the absence of a finite epidemic threshold in the corresponding spreading phenomenon \cite{pastor}.

The degree distribution is only a first way to characterize the degree related topology of a complex network. Indeed, by analyzing scientific collaboration networks, researchers working with many others (high degree nodes) tend to collaborate with other ``hubs''. This means that there exist {\it degree correlations}, and the just mentioned property has been called {\it assortative mixing}, holding generally for social networks \cite{newman1}. On the other hand, in the Internet (at the autonomous system level), high degree nodes are more likely connected to low degree ones, thus exhibiting {\it disassortative} mixing. The influence of such degree correlations on the spreading of an epidemic was investigated in detail \cite{boguna1}, finding that neither assortative nor disassortative scale-free networks exhibit a finite epidemic threshold \cite{boguna2}. The insights about the role of these degree related topological properties in epidemic spreading have been gained at the mean-field level.

Besides degree correlations, triangles are ubiquitous in complex networks as outlined in the first paragraph, and more generally, many loops of short length were found in these systems \cite{caldarelli}. Motivated by Watts and Strogatz' model \cite{watts} which uses a regular lattice, i.e.\ an ordered network possessing many loops, as starting point, we shall also use the concept of {\it local ordering} when referring to the loop structure of a complex network. More information can be extracted if the triangles are sorted according to the degrees of their corners, finding that mainly the low degree nodes account for the high level of clustering \cite{vazquez,ravasz}. This suggests the presence of interesting modular organizations, and similar results were obtained from an analysis of loops of length 4 \cite{caldarelli}. Local ordering plays a crucial role in the function of a metabolic network (with scale-free topology). Indeed, the loop structure is much richer than in the scale-free model based on growth and preferential attachment \cite{gleiss}. This work also investigates the number of triangles as a function of the system size for this model, a result which was generalized to loops up to length 5, yielding robust scaling relations \cite{bianconi}.

Obviously the presence of loops has an effect on the spreading behavior since, with respect to a treelike topology, there exist many more paths along which the virus can propagate. Different strategies in order to gain insights about the role of local ordering properties have been proposed. An interpretation of how clustering influences the stationary spreading behavior was obtained by mapping the epidemic process onto bond percolation \cite{newman2}. Another approach is to abandon the mean-field level and take into account spatial correlations which govern the epidemic dynamics. Matsuda et al. first used the ordinary pair approximation in order to study a population dynamical problem \cite{matsuda}. This approximation, as its name anticipates, accounts for pair correlations and lies at the basis of improved pair models \cite{vanBaalen,keeling} which uncover the role of local ordering in a rather indirect way: clustering enters by making a number of assumptions about the open ($\angle$) and closed ($\triangle$) triple correlations. In cluster approximations, a time-dependent probability is assigned to each configuration of the fundamental cluster whose choice is guided by the network topology \cite{petermann}: for investigating the spreading dynamics on a triangular lattice one uses a triangle as fundamental cluster whereas the star is the appropriate choice for a random network. Higher-order correlations are therefore embedded very explicitly. Moreover the systematic improvability of this method makes it a powerful tool to study probabilistic systems.

In order to understand the role of local ordering properties, the exploration of temporal correlations seems to be an even more natural approach: for example within a two-step description, it matters whether the local topology is treelike or if loops of short length are present. The method is illustrated for the susceptible-infected-susceptible model (see for example \cite{diekmann}), homogeneous networks are used as starting point, and analytical estimates are obtained also for disordered graphs obeying $P(k)=\delta_{k,K}$, $K$ being arbitrary.

The paper is organized as follows. Section II describes the adopted model consisting of the contact network as well as the local dynamics. In section III, we introduce the formalism, from which the two-step description is derived. For completeness, the necessary ingredients in order to arrive at the one-step site approximation, i.e.\ the common mean-field level, are also shown. Section IV explores the implications of the double-step approach for networks of degree 4, the generalization to arbitrary degree is done in the following section. The major conclusions are drawn in section VI.

\section{The Model Ingredients}

The dynamical laws that describe the spreading of an infectious disease are determined by the contact structure which underlies the population. We therefore model the epidemic as a dynamical process on top of a given network that does not change in time. The nodes of the network represent individuals, and the links correspond to relationships between individuals along which an infective agent can propagate.

Since the aim of this paper is the investigation of the role of loops of short length, we adopt a rather simple epidemiological model where the individuals can be only in two possible states, namely infected (I) or susceptible (S). Because the nodes repeatedly run through the cycle susceptible $\rightarrow$ infected $\rightarrow$ susceptible, it is called SIS model. In the physics community, it has recently been formulated as follows \cite{pastor}: A node susceptible to the disease gets infected with probability $\nu \Delta t$ if it is connected to at least one infected nearest neighbor. On the other hand, infected nodes recover spontaneously with probability $\delta \Delta t$. This version of the SIS model is formally advantageous with respect to its conventional formulation, where infected nodes can infect neighboring susceptible vertices with probability $\nu \Delta t$ \cite{diekmann}. In the latter case, susceptible nodes become infected with probability $1-(1-\nu \Delta t)^{k_\text{inf}}$, $k_\text{inf}$ being the number of infected nearest neighbors. In this paper, we will use the former version. By rescaling the time unit, we can reduce the number of parameters to one: the time evolution is determined by the effective spreading rate $\lambda \equiv \nu/\delta$, and the recovery rate is set to 1. The quantitative details of the behavior of the system still depend on the choice of $\Delta t$. In particular, the effect of the loops are of higher order in $\Delta t$, such that their influence is not seen in the continuous-time limit ($\Delta t \rightarrow 0$). As long as $\Delta t > 0$, we set this quantity to 1 without lack of generality.

\begin{figure*}[t]
\includegraphics[width=18cm]{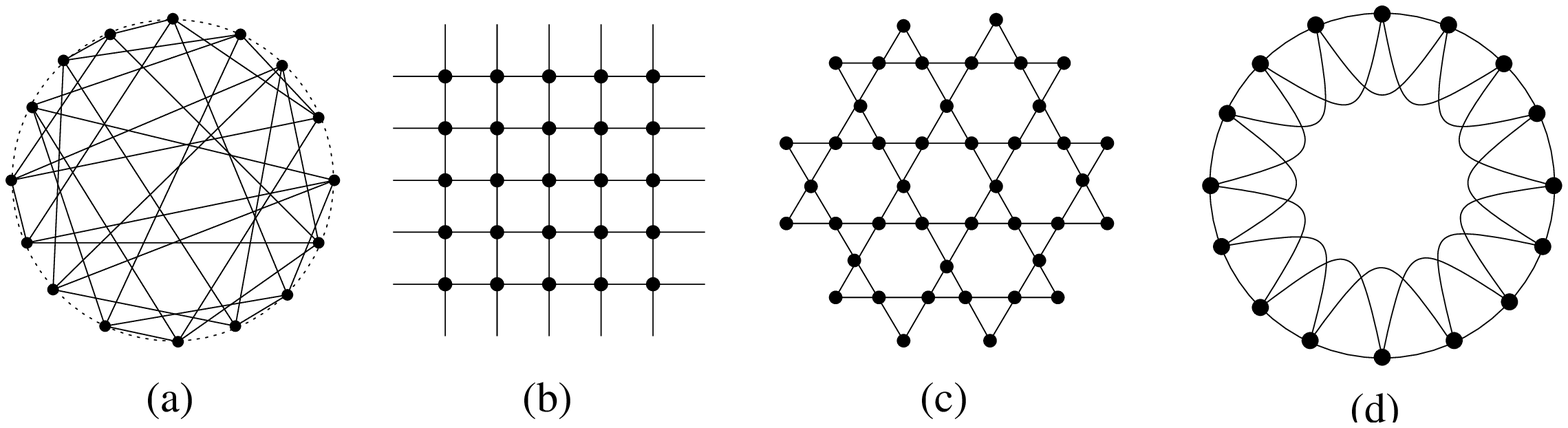}
\caption{Homogeneous networks of degree 4. In (a) the nodes are connected at random under the restriction that $K=4$ links emanate from every vertex, leading to a treelike topology. The role of triangles and loops of length four is studied by means of the square lattice (b) and the Kagom\'e lattice (c) where they appear separately, as well as the ring-type network (d).}
\label{fig:hom4_ex}
\end{figure*}

The other model constituent concerns the underlying network. We shall not attempt to examine the combined effect of the degree distribution, degree correlations and the loop structure. While the role of these degree related connectivity patterns in epidemic spreading is rather well understood, little attention has been given to the effect of local ordering. That is why we focus on this topological property. We therefore use networks in which every node has the same number of nearest neighbors (fixed degree). The adopted strategy is to start with strictly homogeneous graphs (networks in which identical connectivity patterns are ``seen'' from every node) differing in the detailed loop structure (Fig.\ \ref{fig:hom4_ex}). This leads us to an understanding of the role of these distinct local ordering properties, even for networks that are nolonger strictly homogeneous but which still obey $P(k)=\delta_{k,K}$, K being the constant degree.

In the case $K=4$, examples of purely homogeneous networks are the random homogeneous network (Fig.\ \ref{fig:hom4_ex}a) constructed according to the Molloy Reed algorithm \cite{molloy1,molloy2}, the square lattice (Fig.\ \ref{fig:hom4_ex}b), the Kagom\'e lattice (usually used in condensed matter physics as this geometry represents one of the most frustrated antiferromagnetic systems, Fig.\ \ref{fig:hom4_ex}c) and a ring where nodes two units apart from each other are also directly connected (Fig.\ \ref{fig:hom4_ex}d). For the random network, the homogeneity definition given above only holds approximately. While loops of short length do not occur in the limit of large network size, long ones of various lengths may exist. Therefore different nodes may not ``see'' identical connectiviy patterns. Also the Cayley tree lacks completely in loops. But as the last generation of nodes has degree 1, it does not fall into the class of homogeneous networks. Indeed, since the number of nodes belonging to generation $N$ is $K(K-1)^{N-1}$, most of the nodes are even comprised within the last generation. Clearly, all the networks depicted in Fig.\ \ref{fig:hom4_ex} are characterized by the degree distribution $P(k)=\delta_{k4}$, but they differ in the way the second neighbors are arranged, that is the loop structure becomes richer going from (a) to (d). While the random network has no short loops at all, the ones of length 4 are a fingerprint of the square lattice. In the Kagom\'e lattice, every node is a corner of two distinct triangles. Finally, in the ring of degree 4, triangles and two different kinds of loops of length 4 are found, namely plaquettes (as in the square lattice) and the so-called primary quadrilaterals (two adjacent triangles), see section IVB. 

\begin{figure}[h]
\includegraphics[width=8cm]{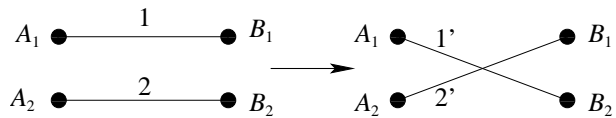}
\caption{Rewiring procedure not affecting the degree distribution: the end vertices of two arbitrarily chosen links are exchanged.}
\label{fig:rewir}
\end{figure}

In order to study the epidemic process on disordered ``homogeneous'' topologies, we subject the networks in Fig.\ \ref{fig:hom4_ex}b-d to the following rearrangement algorithm (Fig.\ \ref{fig:rewir}) \cite{maslov}:
\begin{itemize}
\item{Choose randomly two links (link 1 connecting node $A_1$ with $B_1$ and connection 2 linking vertex $A_2$ with $B_2$) that do not share a common node.}
\item{Remove these two links and establish two new connections between $A_1$ and $B_2$ as well as $A_2$ and $B_1$.}
\end{itemize}

Repeating this rewiring procedure a certain number of times leads to locally varying numbers of loops while the degree distribution remains unaltered.

\section{The Formalism}

In this section, we shall introduce the formalism that is used in order to describe the dynamics on top of the networks specified in the previous section. For completeness, we then derive the (single-step) mean-field approximation which could also be obtained heuristically. In the last paragraph, we derive a two-step description that allows us to gain insight about the loop structure in epidemic spreading.

On an exact level, we shall describe the epidemic dynamics by assigning a probabilty $\mathcal P_t({\bf x})$ to each configuration ${\bf x}$ at every instant of time $t$. The vector ${\bf x}$ contains the states $x_i$ of all the nodes $i$ of the network, $x_i$ being either 0 (susceptible) or 1 (infected). The system probabilities satisfy at any time $t$
$$
\sum_{\bf x} \mathcal P_t({\bf x})=1.
$$
and evolve in time according to
\begin{equation}	\label{equ:exact1}
\mathcal P_{t+1}({\bf x}) =\sum_{\bf y} \mathcal W_{{\bf y} \rightarrow {\bf x}} \mathcal P_t({\bf y}).
\end{equation}
The transition matrix of the system $\mathcal W_{{\bf y} \rightarrow {\bf x}}$ is obtained from the matrix $W_{y_l \rightarrow x_l}^l$ which shall denote the probability that the state of the arbitrary site $l$ changes from $y_l$ to $x_l$, through
$$
\mathcal W_{{\bf y} \rightarrow {\bf x}} = \prod_{l=1}^N W_{y_l \rightarrow x_l}^l,
$$
N being the total number of nodes in the network. The matrix elements representing the probabilites for the possible events at the site $l$ are given by our version of the SIS model, namely
\begin{equation*}	\begin{split}
W_{1 \rightarrow 0}^l &=1 \qquad \qquad W_{0 \rightarrow 1}^l =\lambda\Bigl[1-\prod_{j \text{nn} l} (1-y_j)\Bigr] \\
W_{1 \rightarrow 1}^l &=0 \qquad W_{0 \rightarrow 0}^l =1-\lambda\Bigl[1-\prod_{j \text{nn} l} (1-y_j)\Bigr],
\end{split}	\end{equation*}
or in a more compact form
\begin{equation*}
W_{y_l \rightarrow x_l}^l = 1-x_l+\lambda(2x_l-1)(1-y_l)\Bigl[1-\prod_{j \text{nn} l} (1-y_j)\Bigr].
\end{equation*} 
The products in the above expressions have to be taken over all the nearest neighbors $j$ of node $l$. The factor $1-\prod_{j \text{nn} l} (1-y_j)$ is 1 if at least one $y_j=1$ and 0 otherwise.

Before deriving our two-timestep description, we show how the conventional mean-field approximation is retrieved through this formalism. At that level, the sites are considered independently from each other, and we write for the system probability
\begin{equation}	\label{equ:hom_mix}
\mathcal P_t({\bf x})=\prod_{l=1}^N P_t(x_l),
\end{equation}
i.e.\ the system is described by the single variable $P_t(1)$ [the probability of being susceptible is $P_t(0)=1-P_t(1)$]. Its dynamics is obtained from Eq.\ (\ref{equ:exact1}) by summing it over all possible configurations ${\bf x}$, $x_0$ held fixed
\begin{equation*}
\sum_{\{x_j\}_{j \neq 0}} \mathcal P_{t+1}({\bf x})=\sum_{\bf y} \mathcal P_t({\bf y}) \underbrace{\sum_{\{x_j\}_{j \neq 0}} \mathcal W_{{\bf y} \rightarrow {\bf x}}}_{W_{y_0 \rightarrow x_0}^0},
\end{equation*}
where the node $0$ can be chosen in an arbitrary way. The left hand side of the above equation corresponds to the probability that node $0$ is in state $x_0$ at time $t+1$. With the ansatz (\ref{equ:hom_mix}), the time evolution is
\begin{equation*}
P_{t+1}(1)=\lambda P_t(0)[1-P_t(0)^K].
\end{equation*}
With $P_t(1) \equiv P_t$ and $P_t(0)=1-P_t$, by consequence, we obtain for small values of $P_t$
\begin{equation}	\label{equ:mf_1step}
P_{t+1}=\lambda K (1-P_t)P_t.
\end{equation}
The stationary-state condition ($P_{t+1}=P_t$) leads to a value for the epidemic threshold
\begin{equation}	\label{equ:1step_thr}
\lambda_c=\frac{1}{K}.
\end{equation}
Therefore all the networks depicted in Fig.\ \ref{fig:hom4_ex} are treated identically at this level of description. The only topological property that determines the critical value of the effective spreading rate $\lambda$ is the number of nearest neighbors $K$.

Improvements upon the mean-field description can be obtained by taking into account spatial correlations. Based on the ordinary pair approximation \cite{matsuda}, the presence of triangles enters by establishing a number of hypotheses about the open and closed triple correlations \cite{vanBaalen,keeling}. This approach therefore embeds this topological property in a rather implicit way. In cluster approximations \cite{petermann}, the local topology and its associated spatial correlations translate directly into the choice of the cluster and a set of probabilities for its possible configurations.

Another strategy that serves to incorporate local ordering properties is to take into account temporal correlations. Thus by performing two timesteps exactly, we expect that the way the second neighbors are arranged, enters very naturally into the description. For example, the cases where two nearest neighbors of an arbitrary node are also directly connected (presence of a triangle), where they are linked via a second neighbor (giving rise to a loop of length 4) or where the only path goes through the original node (treelike structure), lead to different results. In the remaining part of this section, we derive the general equation, special cases are then looked at within the following section.

As outlined above, we iterate Eq.\ (\ref{equ:exact1}) once
$$
\mathcal P_{t+1}({\bf x}) =\sum_{\bf y}\Bigl[ \mathcal W_{{\bf y} \rightarrow {\bf x}} \sum_{\bf z} \mathcal W_{{\bf z} \rightarrow {\bf y}} \mathcal P_{t-1}({\bf z})\Bigr]. 
$$
We now again pass to a site approximation. By summing the above equation over all possible configurations ${\bf x}$, $x_0$ held fixed, we get
\begin{equation}	\label{equ:2steps}	\begin{split}
P_{t+1}(x_0)= \sum_{\bf z} \Bigl[ \mathcal P_{t-1}({\bf z})
\sum_{\{y_l\}_{l=0,1,...,K}} \Bigl(W_{y_0 \rightarrow x_0} \prod_{j=0}^K W_{z_j \rightarrow y_j}\Bigr)\Bigr],
\end{split}	\end{equation}
\begin{figure*}[t]
\includegraphics[width=14cm]{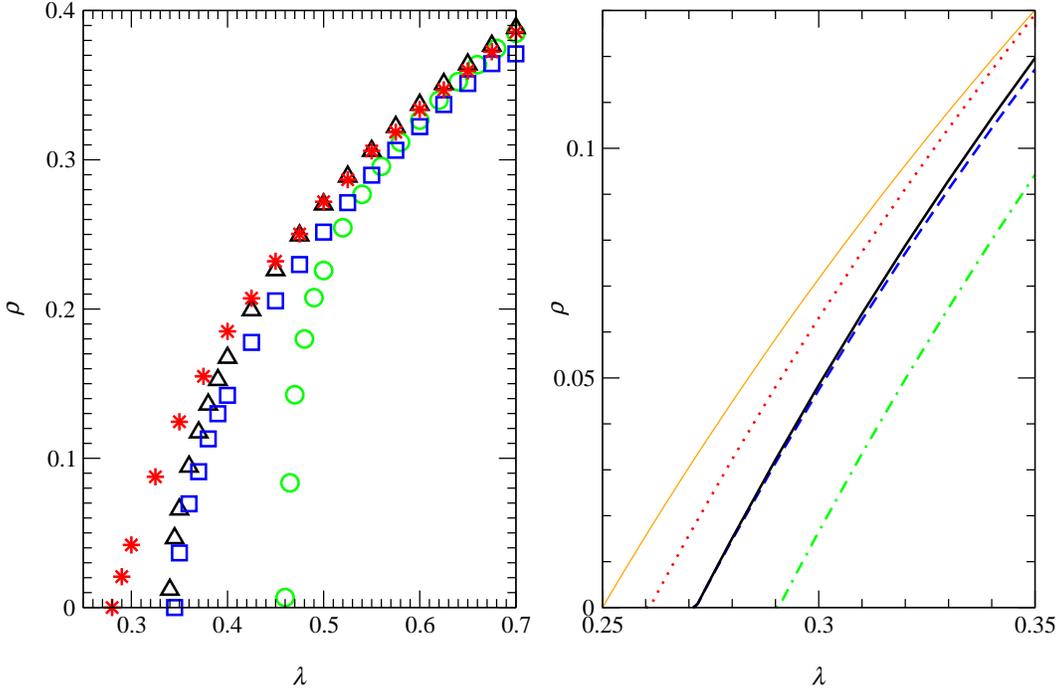}
\caption{Equilibrium prevalences for the epidemic process on top of different networks characterized by $P(k)=\delta_{k4}$. Left: Simulation results for the homogeneous random graph ($\star$), the Kagom\'e lattice ($\triangle$), the square lattice ($\square$) and the ring-type network ($\circ$). Right: The one-step site approximation (thin solid) ignores (local) ordering properties and yields $\lambda_c=0.25$ for any homogeneous network of degree 4. For the Molloy Reed network (dotted), the Kagom\'e lattice (solid), the square lattice (dashed) and the ring (dotted-dashed), different steady-state prevalences are obtained at the two-step level.}
\label{fig:hom4_reg}
\end{figure*}
0 again being an arbitrarily chosen node. Furthermore, the system probability $\mathcal P_{t-1}({\bf z})$ is given by Eq.\ (\ref{equ:hom_mix}), and the nodes $1,2,..,K$ denote the nearest neighbors of the arbitrarily chosen node 0. As only $z$-states associated to the vertex $0$, its nearest and second neighbors appear in the $W$-factors, the sum over the $z$-variables associated to nodes more than 2 links away from vertex 0, is carried out trivially. A tour de force calculation leads to
\begin{widetext}
\begin{multline}		
P_{t+1}(1)=\lambda \Bigl[\lambda \sum_{\alpha_1=1}^K \langle f_{\alpha_1}\rangle_{t-1}
-\lambda^2\Bigl(\sum_{\alpha_1=1}^K \sum_{\alpha_2=\alpha_1+1}^K \langle f_{\alpha_1} f_{\alpha_2}\rangle_{t-1} + \sum_{\alpha_1=1}^K \langle f_0f_{\alpha_1}\rangle_{t-1} \Bigr) \\
+\lambda^3\Bigl(\sum_{\alpha_1=1}^K \sum_{\alpha_2=\alpha_1+1}^K \sum_{\alpha_3=\alpha_2+1}^K \langle f_{\alpha_1} f_{\alpha_2} f_{\alpha_3}\rangle_{t-1} 
+\sum_{\alpha_1=1}^K \sum_{\alpha_2=\alpha_1+1}^K \langle f_0f_{\alpha_1} f_{\alpha_2}\rangle_{t-1}\Bigr)
+ ... \\
-(-\lambda)^K \Bigl(\sum_{\alpha_1=1}^K \sum_{\alpha_2=\alpha_1+1}^K ... \sum_{\alpha_K=\alpha_{K-1}+1}^K \langle f_{\alpha_1} f_{\alpha_2}...f_{\alpha_K}\rangle_{t-1}
+ \sum_{\alpha_1=1}^K \sum_{\alpha_2=\alpha_1+1}^K ... \sum_{\alpha_{K-1}=\alpha_{K-2}+1}^K \langle f_0f_{\alpha_1} f_{\alpha_2}...f_{\alpha_{K-1}}\rangle_{t-1} \Bigr)\Bigr].
\label{equ:exact2}
\end{multline}	
\end{widetext}
\begin{figure*}[t]
\includegraphics[width=17cm]{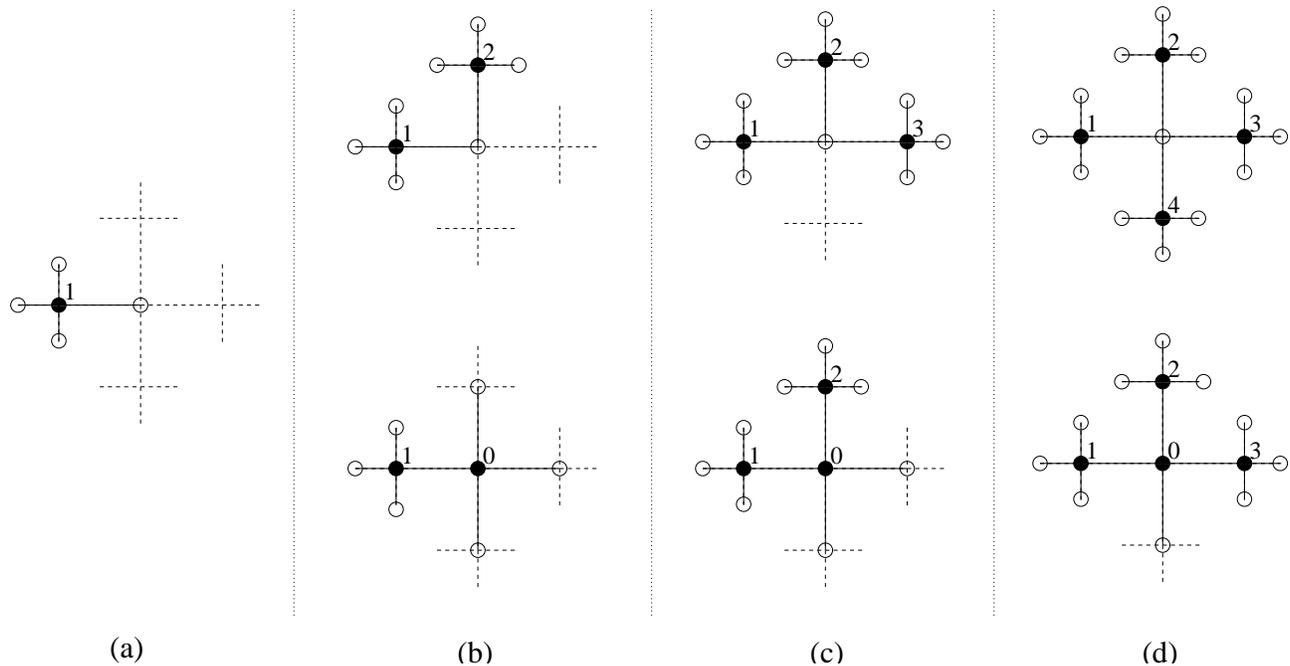}
\caption{Pictorial representation of the terms of Eq.\ (\ref{equ:exact2}) for a treelike topology. The order of a specific subgraph is given by the number of filled circles, (a) corresponding to $\langle f_1\rangle$, (b) to $\langle f_1f_2\rangle$ and $\langle f_0f_1\rangle$, and so forth. In this vein, the indices of the $f$-factors appearing in the expectation values correspond to filled circles whereas empty circles represent their nearest neighbors. The dashed lines are the links of the complete graph which are not contained in a specific subgraph.}
\label{fig:tree_sub}
\end{figure*}
Thereby the connectivity embedding factor
\begin{equation*}	\begin{split}
f_{\alpha}&\equiv (1-z_\alpha)\Bigl[1-\prod_{\sigma \text{nn} \alpha} (1-z_\sigma)\Bigr] \\
&=\begin{cases}
1& \text{if $z_\alpha=0$ and at least one $z_\sigma=1$,}\\
0& \text{otherwise.}
\end{cases}
\end{split}	\end{equation*}
and the expectation value of a function of the states of node 0, its nearest and second neighbors, these vertices collectively being denoted by $\mathcal N^2$,
\begin{equation}	\label{equ:expect}
{\Big \langle} g(\{z_k\}_{k \in \mathcal N^2}){\Big \rangle}_t \equiv \sum_{\{z_k\}_{k \in \mathcal N^2}} \Bigl[ \prod_{l \in \mathcal N^2} P_t(z_l) g(\{z_m\}_{m \in \mathcal N^2})\Bigr]
\end{equation}
were introduced for notational convenience. In the following, an expectation value of a product of $n$ $f$-factors will be referred to as a term of $n$-th order although it is proportional to $\lambda\cdot\lambda^n$. As is illustrated in detail in the next section, every term of Eq.\ (\ref{equ:exact2}) corresponds to a subgraph of the graph composed of the nodes $\mathcal N^2$ and whose links are according to the network under investigation. It can already be anticipated that the first term accounts for the degree distribution only, whereas the contributions of higher order will give insight about the role of the loop structure.

\section{Networks of degree 4}

In this section, we elaborate the implications of our two-step description, the analysis being restricted to the stationary state. 

In the left part of Fig.\ \ref{fig:hom4_reg}, we show the simulation results for the graphs introduced in section II. The ring-type network exhibits the largest epidemic threshold. For both the square and Kagom\'e lattices, the critical value is $\lambda_c \simeq 0.34$. We therefore anticipate that either four plaquettes or two triangles (per node) lead to the same effect in the regime of low prevalences. Finally, the lowest epidemic threshold is found if the population is arranged on a homogeneous random network (of degree 4). The last result is very intuitive since in such a graph, global distances are small, making it more easy for a virus to spread. Therefore, even if the effective spreading rate is rather low, a finite fraction of the population will be infected in the stationary state, hence the small value for the location of the onset of the epidemic. In summary, these results indicate that the poorer the loop structure, the lower the corresponding epidemic threshold. 

In the right part of Fig.\ \ref{fig:hom4_reg}, the one-step and two-step site approximations are reported. The former corresponds to the steady-state solution of Eq.\ (\ref{equ:mf_1step}) for which $\rho=0$ at $\lambda_c=1/4$ according to Eq.\ (\ref{equ:1step_thr}). All the networks in question are therefore treated identically, the loop structure being ignored at this level of description. Yet, the two-step solutions [Eq.\ (\ref{equ:2steps})] are diverse for the different graphs. Going from right to left, the curves correspond to the ring, the square lattice, the Kagom\'e lattice and the Molloy-Reed network, that is they appear in the same sequence as at the level of simulation. Furthermore, the curves corresponding to the Kagom\'e and square lattice also meet the $x$-axis at the same value of $\lambda$. These findings confirm our intuitive arguments given in the previous section. It has to be noted that the two-step estimates for the threshold values are still considerably inaccurate especially for the ring and lattices, but this just highlights the presence of higher-order spatiotemporal correlations. However, the important point is that the degeneracy associated to the one-step description disappers at the two-step level. 

On the basis of Eq.\ (\ref{equ:exact2}), we shall now analytically study the effect of local ordering properties upon the epidemic spreading, leading to a quantitative understanding of the threshold value.

\subsection{Random network}

We shall now evaluate all the terms of Eq.\ (\ref{equ:exact2}) for a locally treelike topology. Fig.\ \ref{fig:tree_sub} shows the subgraphs representing the terms in Eq.\ (\ref{equ:exact2}), in increasing order. Thereby the correspondance is as follows: Given the term $\langle f_\alpha f_\beta \rangle$, the nodes $\alpha$ and $\beta$ are represented by filled circles whereas their nearest neighbors are drawn by empty circles. The links which enter at the level of the subgraph in question, are represented by solid lines whereas the ignored ones are dashed. If we denote the second neighbors of the central vertex 0 by $l1, l2,l3$ for $l=1,2,3,4$ and follow Eq.\ (\ref{equ:expect}), the first order contribution for $\alpha_1=1$ (subgraph in Fig.\ \ref{fig:tree_sub}a) is
\begin{equation*}	\begin{split}
\langle f_1 \rangle = &\sum_{z_1}\sum_{z_0}\sum_{z_{11}..z_{13}}\Bigl\{P(z_1)P(z_0)P(z_{11})P(z_{12}) P(z_{13}) \\
&\times \underbrace{(1-z_1)[1-(1-z_0)(1-z_{11})(1-z_{12})(1-z_{13})]}_{f_1}\Bigr\} \\
&\qquad =4P + \mathcal O(P^2)
\end{split}	\end{equation*}
where the sum over the $z$-variables to which no circles are associated, has been carried out trivially. Furthermore, we again have set $P \equiv P(1)$ in the third line (this is also done below), and the time index was omitted since we are only interested in the steady state. This term appears with multiplicity 4 (due to $\sum_{\alpha_1=1}^K$), giving the contribution $16P$ to first order in $P$.

Fig.\ \ref{fig:tree_sub}b shows the subgraphs representing $\langle f_1f_2\rangle$ (upper part) and $\langle f_0f_1\rangle$ (lower part). Their contributions are
\begin{equation*}	\begin{split}
\langle f_1f_2 \rangle=&\sum_{z_0}\sum_{z_1}\sum_{z_2}\sum_{z_{11}..z_{13}}\sum_{z_{21}..z_{23}} \Bigl\{ P(z_0)P(z_1)P(z_2) \\
&\quad \times P(z_{11})P(z_{12})P(z_{13})P(z_{21})P(z_{22})P(z_{23})f_1f_2\Bigr\} \\
&\qquad\qquad = P+ \mathcal O (P^2),
\end{split}	\end{equation*}
occuring ${4 \choose 2}=6$ times and 
\begin{equation*}	\begin{split}
\langle f_0f_1 \rangle =&\sum_{z_0}\sum_{z_1}\sum_{z_2..z_4}\sum_{z_{11}..z_{13}}\Bigl\{P(z_0)P(z_1)P(z_2)P(z_3)P(z_4) \\
&\qquad \times P(z_{11})P(z_{12})P(z_{13})f_0f_1\Bigr\} \\
&\qquad\qquad = 9P^2+\mathcal O (P^3),
\end{split}	\end{equation*}
thus not giving a contribution to first order in $P$. As a consequence, the second-order term (that is the one proportional to $\lambda\cdot\lambda^2$) is $6P$.

As the procedure should now be clear, we only give the results for the remaining orders. The upper subgraph of Fig.\ \ref{fig:tree_sub}c represents the term $\langle f_1f_2f_3\rangle$. Its contribution is $P+\mathcal O (P^2)$. The lower subgraph corresponds to $\langle f_0f_1f_2\rangle$, yielding $18P^3+\mathcal O (P^4)$. As the former term has multiplicity ${4 \choose 3}=4$, the total third-order contribution is $4P$.

As far as the fourth order is concerned (Fig.\ \ref{fig:tree_sub}d), the subgraph involving node 0 as filled circle neither gives a contribution whereas the term $\langle f_1f_2f_3f_4\rangle$ having multiplicity 1 also gives $P+\mathcal O (P^2)$, thus totally yielding $\lambda\cdot\lambda^4 P$.

Collecting these findings, we obtain the following condition that determines the epidemic threshold for a treelike topology
\begin{equation}	\label{equ:tree4_eq}
1=\lambda(16\lambda-6\lambda^2+4\lambda^3-\lambda^4),
\end{equation}
which is satisfied by $\lambda_c \simeq 0.2609$. This is the value that can be extracted from the right panel of Fig.\ \ref{fig:hom4_reg} (second curve from the left).

\subsection{Graphs with loops}

\begin{figure}[t]
\includegraphics[width=7cm]{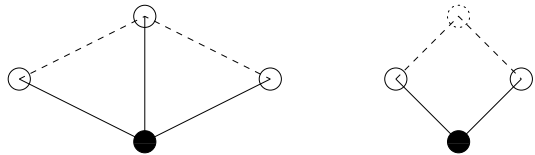}
\caption{Loops of length 4 in a network. Primary quadrilaterals (left) are two adjacent triangles and therefore involve only nearest neighbors (solid empty circles) of the central node (filled circle). A secondary quadrilateral (right) involves a second neighbor (dotted empty circle) as well.}
\label{fig:quadlat}
\end{figure}

It is easy to imagine that the preceeding analysis yields different results when triangles and loops of length 4 are present. Caldarelli et al.\ \cite{caldarelli} have classified loops of length 4 in a complex network into {\it primary} and {\it secondary quadrilaterals} (Fig.\ \ref{fig:quadlat}). In the former case, the external vertices which the loop is composed of are all nearest neighbors whereas secondary quadrilaterals are plaquettes, the external nodes being two nearest and a second neighbor. With these concepts, the loop structure of a strictly homogeneous graph can quantitatively be characterized as follows: By choosing an arbitrary node, the number of edges between its nearest neighbors is denoted by $E$. $Q_1$ and $Q_2$ shall refer to the number of primary and secondary quadrilaterals. For the networks depicted in Fig.\ \ref{fig:hom4_ex}b-d, we report the corresponding values in Tab.\ \ref{tab:simplenets_EQ}.

\begin{table}[h]
\caption{Loop properties for the simple non-treelike networks described in section II.}
\begin{ruledtabular}
\begin{tabular}{cccc}
 & $E$ & $Q_1$ & $Q_2$ \\
\hline
square lattice & 0 & 0 & 4 \\
Kagom\'e lattice & 2 & 0 & 0 \\
ring & 3 & 2 & 2 \\
\end{tabular}
\end{ruledtabular}
\label{tab:simplenets_EQ}
\end{table}

\begin{figure}[t]
\includegraphics[width=7.5cm]{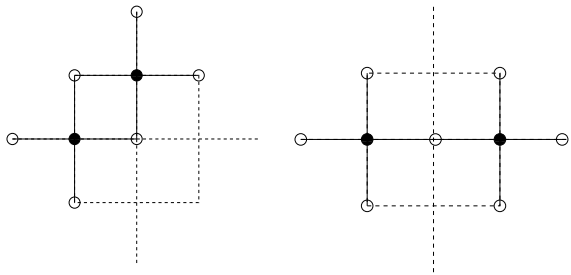}
\caption{Second-order subgraphs not involving the central node as filled circle for the square lattice. To first order in $P$, the left subgraph yields the contribution $2P$, the right one $P$. For further explanations, see Fig.\ \ref{fig:tree_sub}.}
\label{fig:sq2ord}
\end{figure}

Let us now look at the subgraph development for the square lattice whereby we focus on the important changes with respect to the treelike case. The reader interested in the full elaborations is referred to the appendix. We have already noticed that the first-order term is fully determined by the degree distribution, therefore the $\lambda\cdot\lambda$ coefficient is 16, as in the treelike case. At order 2, the term $\langle f_1f_2\rangle$ (upper subgraph of Fig.\ \ref{fig:tree_sub}b) splits into two subgraphs in the presence of plaquettes (Fig.\ \ref{fig:sq2ord}). The right subgraph is the same as in the treelike case, yet the left yields a contribution $2P+\mathcal O(P^2)$. Their multiplicities are 4 (left) and 2 (right) summing up to ${4 \choose 2}=6$. The resulting $\lambda\cdot\lambda^2$ coefficient is therefore -10. Although different subgraphs enter into the development also at the orders $\geq 3$, the coefficients appearing in the equation determining the epidemic threshold do not change.

\begin{figure}[h]
\includegraphics[width=7cm]{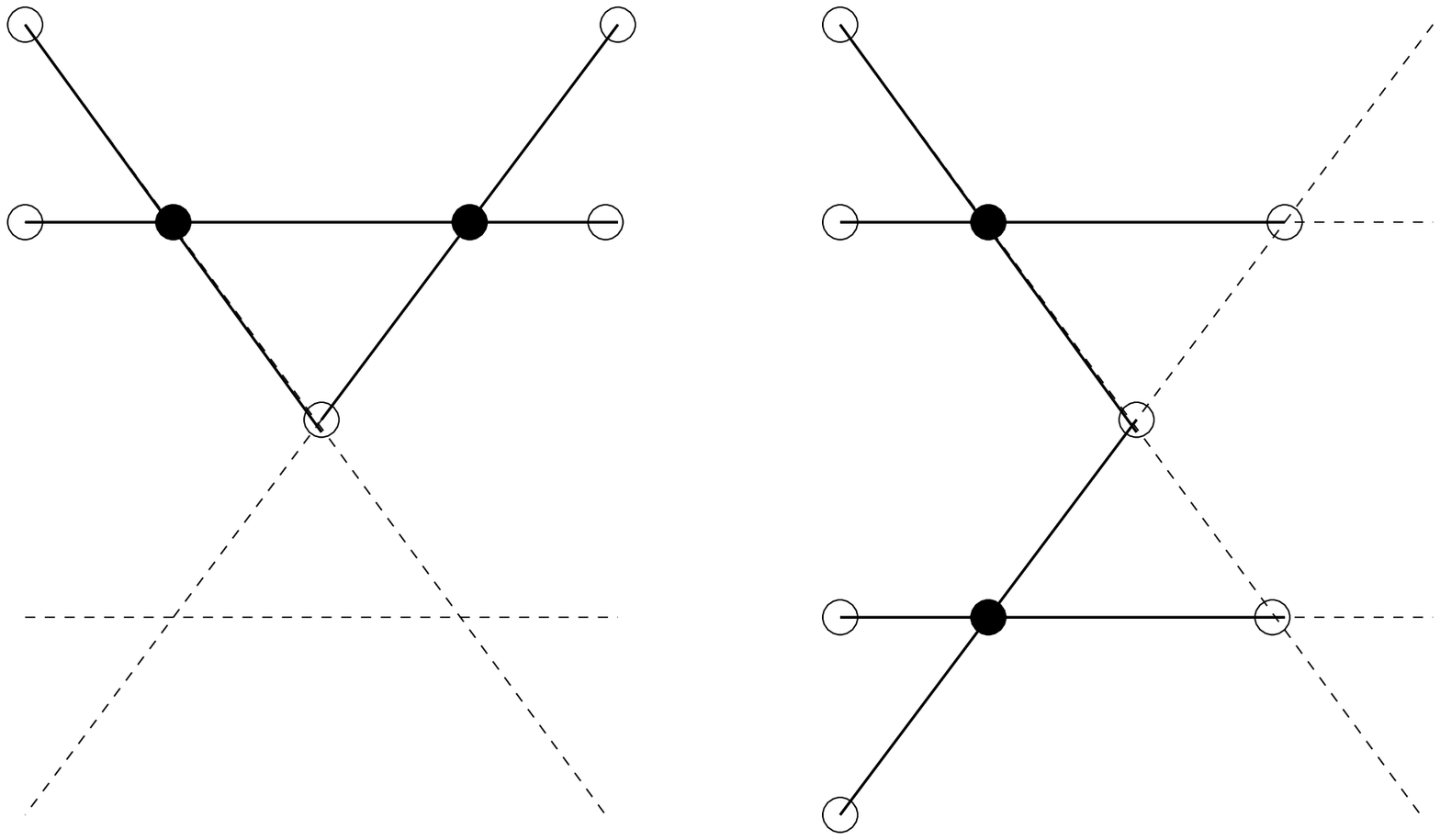}
\caption{Subgraphs of second order for the Kagom\'e lattice, both contribute with $P+\mathcal O(P^2)$. See Fig.\ \ref{fig:tree_sub} for details regarding line- and circle styles.}
\label{fig:kag2ord}
\end{figure}

The second-order subgraphs for the Kagom\'e lattice are depicted in Fig.\ \ref{fig:kag2ord}. Both's contributions are $P+\mathcal O(P^2)$. The one involving a triangle appears 6 times whereas the right subgraph has multiplicity 4. We therefore obtain the same second-order coefficient as for the square lattice. An analysis for the higher-order subgraphs yields no difference with respect to the square lattice. These two cases are therefore equivalent at the two-step level for $P \ll 1$. 

In Tab.\ \ref{tab:simplenets_coeff}, we summarize the coefficients for these two lattices as well as the ring (Fig.\ \ref{fig:hom4_ex}d). The full developments are given in the appendix.

\begin{table}[t]
\caption{Coefficients of the two-step threshold equation for our networks having in common $P(k)=\delta_{k4}$, but differing in the loop pattern.}
\begin{ruledtabular}
\begin{tabular}{ccccc}
$\lambda\cdot\lambda^n$- coeff. & $n=1$ & $n=2$ & $n=3$ & $n=4$ \\
\hline
square lattice & 16 & -10 & 4 & -1 \\
Kagom\'e lattice & 16 & -10 & 4 & -1 \\
ring & 16 & -16 & 6 & -1
\end{tabular}
\end{ruledtabular}
\label{tab:simplenets_coeff}
\end{table}

Of course the idea is now to extend Eq.\ (\ref{equ:tree4_eq}) such that it holds for all the investigated graphs. Our findings suggest that the local ordering properties enter in the following way into the equation determining the epidemic threshold:
\begin{equation}	\label{equ:thres_uptoQ2}
1=\lambda\Bigl[16\lambda -(6+2E+Q_1+Q_2)\lambda^2 \\
+(4+Q_1)\lambda^3-\lambda^4\Bigr].
\end{equation}

\begin{figure}[h]
\includegraphics[width=8cm]{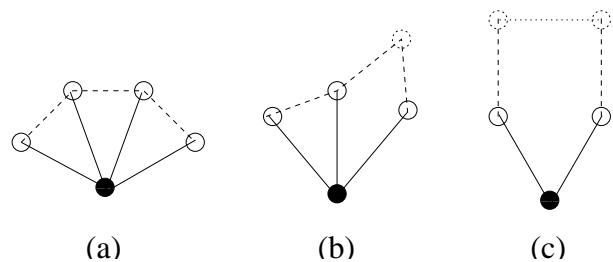}
\caption{Loops of length 5 involving different hierarchies of nearest neighbors. Connections emanating from the central node (filled circle) are drawn as a solid line, links going from nearest neighbors (empty solid circle) to other nearest neighbors or second neighbors (empty dotted circle) are dashed, and second neighbors are connected by a dotted line. The pentalateral (a) involves nearest neighbors only, in (b) the loop traverses a second neighbor and in the one in (c) lacking in internal connections, a link between two second neighbors serves to close it.}
\label{fig:pentalat}
\end{figure}

\begin{figure}[t]
\includegraphics[width=7.5cm]{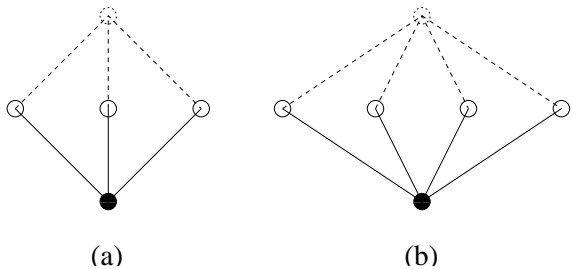}
\caption{Quadrilaterals of orders 3 and 4. The $n$ (a: $n=3$, b: $n=4$) nearest neighbors (empty solid circles) of the central node (filled circle) share another vertex (dotted empty circle) as nearest neighbor.}
\label{fig:quadlat_34}
\end{figure}

However this is not the full story. What about loops of length 5? Let us argue why they do not enter in the framework of a two-step description. Although there exist such loops involving only first and second neighbors (Figs.\ \ref{fig:pentalat}a and b), it may also be closed only between two second neighbors (Fig.\ \ref{fig:pentalat}c). Obviously such a connection is ignored at the two-step level. In the language of graph theory \cite{gross}, the latter case corresponds to a {\it fundamental} loop whereas the former examples can be reduced to loops of length 3 and 4. But whatever the number of hierarchies of nearest neighbors involved in the formation of the loop is, the point is the following. If the central node is infected at time $t$, it can causally affect only vertices two links away, corresponding to a chain of 4 links. Obviously, it matters whether the first and the last node of this chain are identical. In this case, we have a loop of length 4. Otherwise it cannot be distinguished whether the topology is fully treelike of if loops of length greater than 4 are present. Along these lines, it has to be expected that loops up to length $2n$ enter within an $n$-timestep description. In contrast, the presence of higher-order quadrilaterals modifies the coefficients of Eq.\ (\ref{equ:thres_uptoQ2}). Fig.\ \ref{fig:quadlat_34}a shows what we shall call a {\it tertiary} quadrilateral: the three nearest neighbors of the central node are all connected to another common node. Obviously, the presence of a tertiary quadrilateral implies $Q_2={3 \choose 2}=3$ secondary quadrilaterals. In a {\it fourth-order} quadrilateral (Fig.\ \ref{fig:quadlat_34}b), 4 nodes share two common vertices as nearest neighbours, implying the presence of $Q_3={4 \choose 3}=4$ tertiary quadrilaterals and $Q_2={4 \choose 2}=6$ secondary quadrilaterals. In a network of degree $K>4$, quadrilaterals up to order $K$ can in principle be found.

\begin{figure}[h]
\includegraphics[width=6cm]{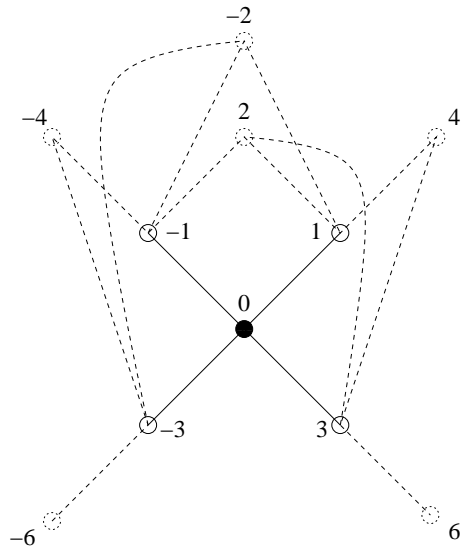}
\caption{This figure visualizes how the nearest (empty solid circles) and second neighbors (empty dotted circles, vertices \{-3,-1,1,3\}) of an arbitrarily chosen node (filled circle, node 0) are arranged in a one-dimensional lattice with additional connections between sites 3 units apart. The sets \{0,\{-1,1,3\},2\} as well as \{0,\{-3,-1,1\},-2\} are forming tertiary quadrilaterals.}
\label{fig:ring2space}
\end{figure}

A one-dimensional lattice with additional connections between the nodes $i$ and $i+3$ for all $i$ (instead of $i+2$ as in the ring investigated up to now, Fig.\ \ref{fig:hom4_ex}d) possesses the neighbourhood structure shown in Fig.\ \ref{fig:ring2space}, i.e.\ it is characterized by $E=Q_1=0, Q_2=8, Q_3=2$ and $Q_4=0$. By applying our formalism to this case and to a network that has fourth-order quadrilaterals, Eq.\ (\ref{equ:thres_uptoQ2}) generalizes to
\begin{equation}	\begin{split}	\label{equ:thr_hom4}
1=\lambda\Bigl[16\lambda &-(6+2E+Q_1+Q_2)\lambda^2 \\
&+(4+Q_1+Q_3)\lambda^3-(1+Q_4)\lambda^4\Bigr],
\end{split}	\end{equation}
the coefficients of order 3 and 4 being modified only.

\subsection{Introducing disorder}

\begin{figure*}[t]
\includegraphics[width=16cm]{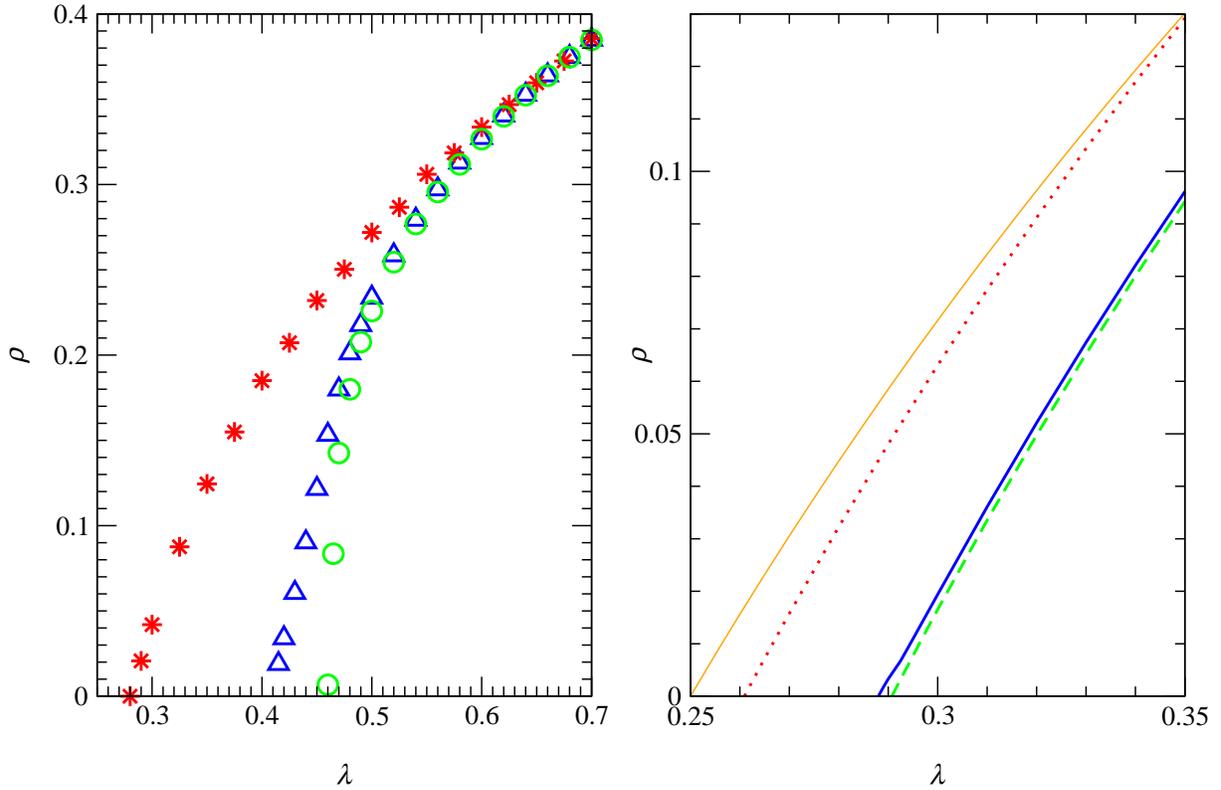}
\caption{Steady-state prevalence for the ring-type network having undergone different degrees of randomization. Left: Simulation result for the original network ($\circ$), its partially rewired version ($\triangle$) and the entirely random network ($\star$). Right: The single-step site approximation (thin solid) treats all the networks in question identically whereas at the two-step level, the Molloy Reed network (dotted), the partially randomized ring (solid) and the fully ordered ring (dashed) appear in the same sequence as in the left panel.}
\label{fig:ring4_100rand}
\end{figure*}

The networks considered up to now lack in the small world phenomenon, a property characterizing social networks on which the epidemic process is occuring. By starting with a ring-like network (Fig.\ \ref{fig:hom4_ex}d) and repeating the rewiring procedure described in section II a certain number of times, we obtain graphs of fixed degree $K=4$ that are simultaneously highly clustered, and in which the average distance between pairs of nodes is small \cite{watts}.

The left part of Fig.\ \ref{fig:ring4_100rand} reports the simulation result for the equilibrium prevalence of the epidemic process on the disordered ring. Systems of size $N=10^4$ were used, and the rewiring procedure was repeated $n=100$ times. For completeness, the two limiting cases (fully ordered ring and random network) are also depicted. This panel shows the considerable effect on the steady-state spreading behavior of the rather small number of rewirings. 

The right panel of this figure depicts the one-step site approximation (predicting $\lambda_c=0.25$ for all cases) and the numerical solutions of the two-step description (\ref{equ:2steps}). It has to be noted that for the partially rewired ring lacking in strict homogeneity, Eq.\ (\ref{equ:2steps}) was solved at every node, therefore involving the set $P_i(x_i)$, $i=1,2,...,N$, the resulting prevalence being given by $1/N \sum_{i=1}^N P_i(1)$. Again at the double-step level, the networks in question are treated differently, as it could already be observed in Fig.\ \ref{fig:hom4_reg}. Here the curve corresponding to the partially randomized ring lies closer to the original network in proportion to the simulation result. This is due to the small world property which can have a considerable effect on the location of the onset of the epidemic. Obviously, the simulation result uncovers the real effect of this global topological property whereas at the two-step level, it is the slightly poorer loop structure that accounts for the corresponding shift in the epidemic threshold.

Of course the quantities $E, Q_1, Q_2, Q_3$ and $Q_4$ are nolonger reasonable for a partially randomized network due to the lack of strict homogeneity, but rather its local ordering properties can be quantified by averaging these values over the entire network. The emerging topological parameters $\bar E$ and $\bar Q_i$ ($i=1,2,3,4$) are essentially the {\it clustering-} (up to the factor $K(K-1)/2=6$) and {\it grid-coefficients}, i.e.\ the densities of triangles and loops of length 4 \cite{watts,caldarelli}. We may therefore replace $E$ and the number of quadrilaterals (of the different orders) in Eq.\ (\ref{equ:thr_hom4}) by its mean-values, yielding the following estimate for the epidemic threshold condition
\begin{equation} 	\label{equ:thr_rand4}	\begin{split}
1=\lambda\Bigl[16\lambda &-(6+2\bar E+\bar Q_1+\bar Q_2)\lambda^2 \\
&+(4+\bar Q_1+\bar Q_3)\lambda^3-(1+\bar Q_4)\lambda^4\Bigr].
\end{split}	\end{equation}
For our partially randomized ring, we have $\bar E=2.883, \bar Q_1=1.886, \bar Q_2=1.958$ and $\bar Q_3=\bar Q_4=0$, leading to $\lambda_c \simeq 0.2892$. This value corresponds approximately to where the corresponding curve in the right part of Fig.\ \ref{fig:ring4_100rand} meets the $x$-axis.

\section{Arbitrary degree}

The implications of our two-step description have been illustrated for homogeneous networks of degree 4 in the previous section. This was a convenient choice as there exists a number of familiar simple graphs obeying $P(k)=\delta_{k4}$, differently ordered. Of course our formalism enables us to generalize the obtained threshold condition (\ref{equ:thr_hom4}) to an arbitrary degree $K$, which is the subject of this section.

Let us again look at a fully treelike network, using Eqs.\ (\ref{equ:exact2}) and (\ref{equ:tree4_eq}) as guidelines. The $\lambda^2$-coefficient 16 incorporating the degree distribution is simply $4\cdot 4$ since $\langle f_\alpha\rangle=4P+\mathcal O(P^2)$ and $\alpha$ runs from 1 to 4. The remaining coefficients -6, 4 and -1 correspond to the binomial coefficients $-{4 \choose 2}, {4 \choose 3}$ and $-{4 \choose 4}$. Indeed the threshold equation for a treelike network of degree $K$ derived by Eq.\ (\ref{equ:exact2}) is
\begin{equation}	\label{equ:tree_gen_eq}
1=\lambda\Bigl[ \lambda K^2-\sum_{\kappa=2}^K \lambda^\kappa {K \choose \kappa}\Bigr].
\end{equation}

Repeating the graph developments for homogeneous networks characterized by different values of $K$ and varying loop structures reveals that the very same correction terms enter into Eq.\ (\ref{equ:tree_gen_eq}), yielding
\begin{equation}	\label{equ:final_hom}	\begin{split}
1= \lambda \Bigl\{ &\Theta(K-1) K^2 \lambda  \\
-&\Theta(K-2)\Bigl[{K \choose 2}+2E+Q_1+Q_2\Bigr]\lambda^2 \\
+&\Theta(K-3)\Bigl[{K \choose 3}+Q_1+Q_3\Bigr]\lambda^3 \\
-\sum_{\kappa=4}^K &\Theta(K-\kappa)\Bigl[{K \choose \kappa}+Q_\kappa\Bigr](-\lambda)^\kappa\Bigr\}
\end{split}	\end{equation}
where $E$ is again the number of connections between the nearest neighbors of an arbitrarily chosen node, $Q_n$ denotes the number of quadrilaterals of order $n$ and $\Theta(x)$ is the step-function defined by
$$
\Theta(x)=\begin{cases}
1& \text{if } x \ge 0\\
0& \text{otherwise.}
\end{cases} 
$$
The equation that describes the role of local ordering in a disordered network subject to $P(k)=\delta_{k,K}$ is again obtained simply by replacing the corresponding quantities by  its mean-values ($E \rightarrow \bar E$, $Q_i \rightarrow \bar Q_i$) in Eq.\ (\ref{equ:final_hom}), providing an improved estimate for the epidemic threshold.

\section{Conclusion}

The spreading of an infectious disease was modeled as a dynamical process on top of a contact network. We used a discrete-time version of the simple SIS-model, that is infected nodes recover with probability $\Delta t$, and susceptible nodes become infected with probability $\lambda \Delta t$ if they are connected to at least one infected nearest neighbor. As far as the connectivity patterns underlying the population are concerned, we chose homogeneous networks as starting point and introduced an arbitrary degree of disorder by an appropriate rewiring procedure not affecting the degree distribution $P(k)=\delta_{k,K}$.

Describing the epidemic dynamics of the entire population as a Markovian process, we derived a two-step description that takes into account temporal correlations. This approach revealed to be very prolific if one wants to unravel the role of loops of short length in the contact network regarding epidemic spreading. Indeed it leads to a subgraph development where the complete graph involves the connectivity patterns of two hierarchies of nearest neighbors (of an arbitrarily chosen node). Within this novel approach serving to tract a probabilistic system, the local topology, be it treelike or be loops of length 3 or 4 present, therefore enters very naturally. The analytically obtained condition for the location of the onset of the epidemic then serves as a guiding equation elucidating the role of clustering and grid-like ordering in epidemic spreading.

In principle it is possible to apply our two-step description to more complex networks where different degrees are present, uncovering the effect of the degree-dependent densities of triangles and loops quadrilaterals on the critical value. Likewise, loops of length up to $2n$ are expected to enter within an $n$-timestep description, also providing a natural classification of them. However the major insight gained by the strategy of exploring temporal correlations is best illustrated as it was done in this paper.

\section{Acknowledgment}

We wish to thank EC-Fet Open project COSIN IST-2001-33555, and the OFES-Bern (CH) for financial support.

\appendix*
\section{Full subgraph developments}

Here we show the complete subgraph developments for the square lattice (Tab.\ \ref{tab:sq_sub}), the Kagom\'e lattice (Tab.\ \ref{tab:kag_sub}) and the ring-type network of Fig.\ \ref{fig:hom4_ex}d (Tab.\ \ref{tab:ring_sub}). Every subgraph corresponds to a term in Eq.\ (\ref{equ:exact2}), its contribution is obtained by the procedure illustrated in section IVA. The $\lambda\cdot\lambda^n$-coefficient of the threshold equation is obtained by summing all the $\mathcal O(P)$ contributions (taking into account the multiplicities) of the $n$-th-order subgraphs.

\begin{table*}[h]
\caption{Subgraph development for the square lattice. All the terms in Eq.\ (\ref{equ:exact2}) are symbolized by a specific subgraph, its order being given by the number of filled circles. The $\lambda\cdot\lambda^2$-coefficient -10, as an example, is obtained by summing the various $\mathcal O(P)$ contributions, that is $2\cdot 4+1\cdot 2+0\cdot 4=10$, and the negative sign comes from Eq.\ (\ref{equ:exact2}). The $\lambda\cdot\lambda^3$- and $\lambda\cdot\lambda^4$-coefficients are unaltered with respect to the treelike topology although other subgraphs enter into the development.}
\begin{ruledtabular}	
\begin{tabular}{|c|c|c|c|c|}
order ($n$) & 1 & 2 & 3 & 4 \\
\hline
subgraph & & & & \\
 & \includegraphics[width=3.5cm]{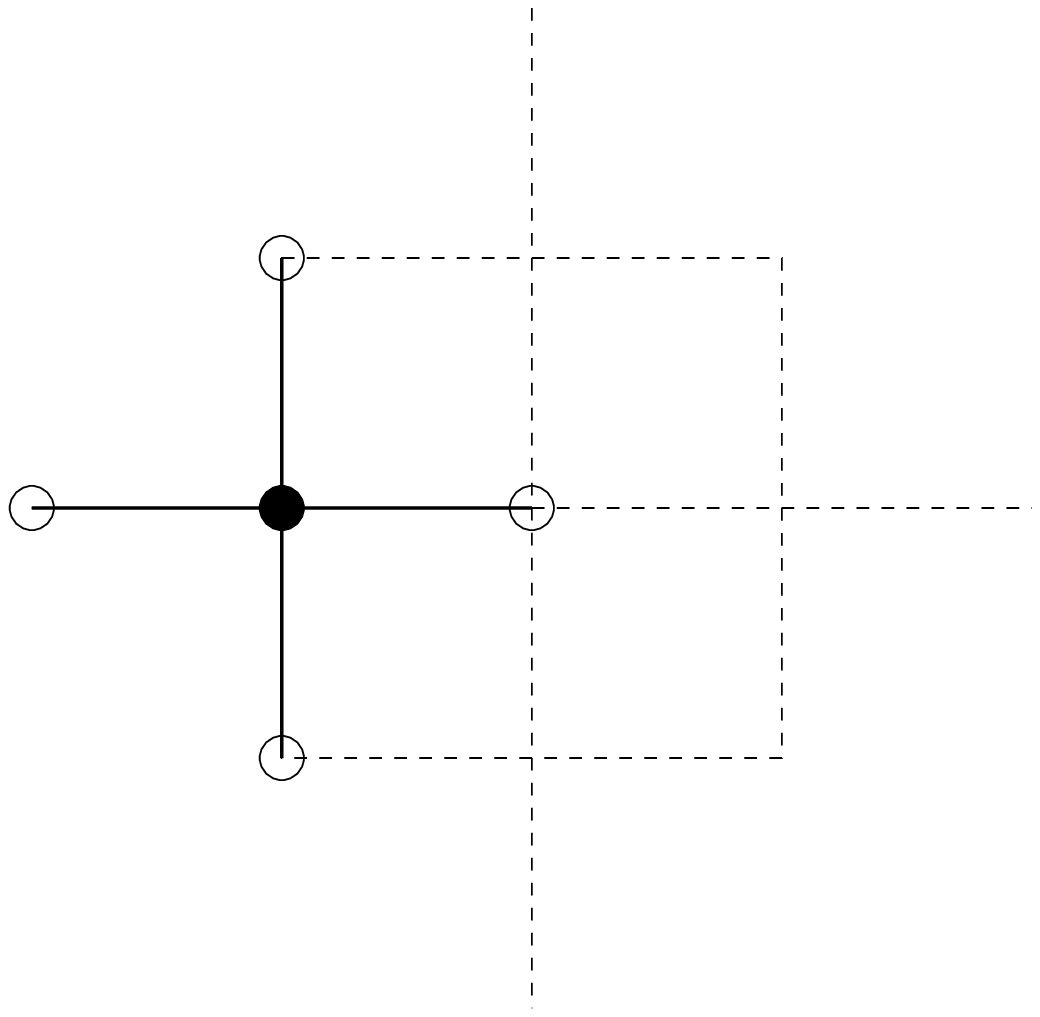} & \includegraphics[width=3.5cm]{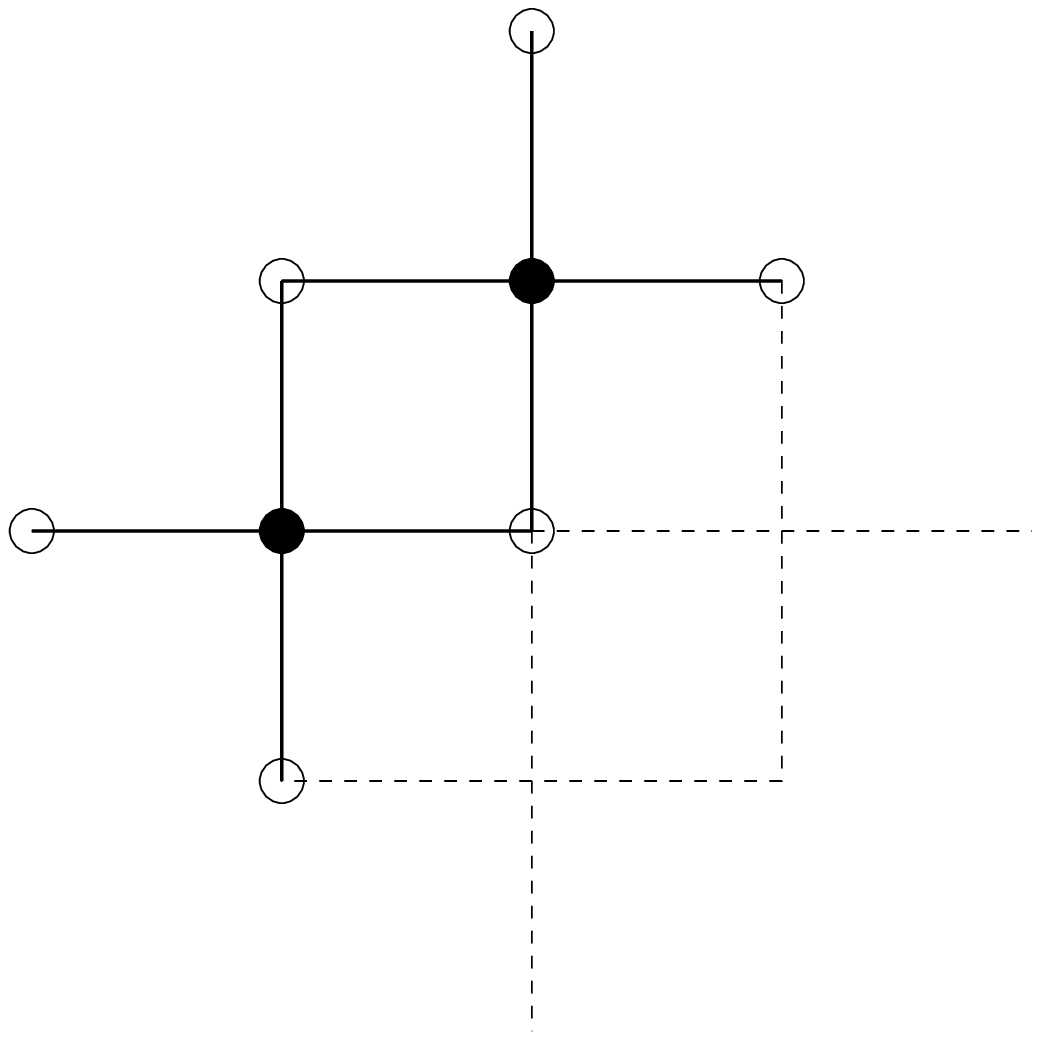} & \includegraphics[width=3.5cm]{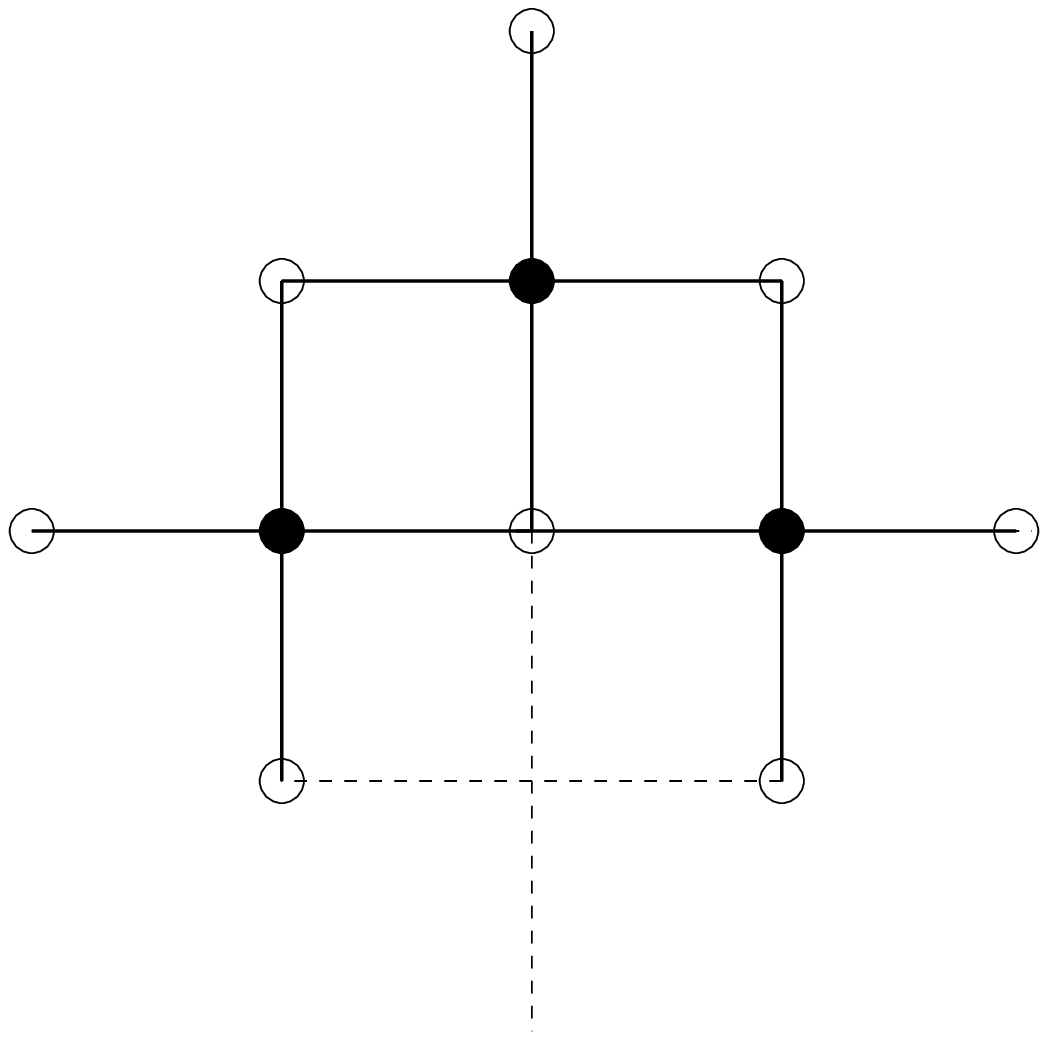} & \includegraphics[width=3.5cm]{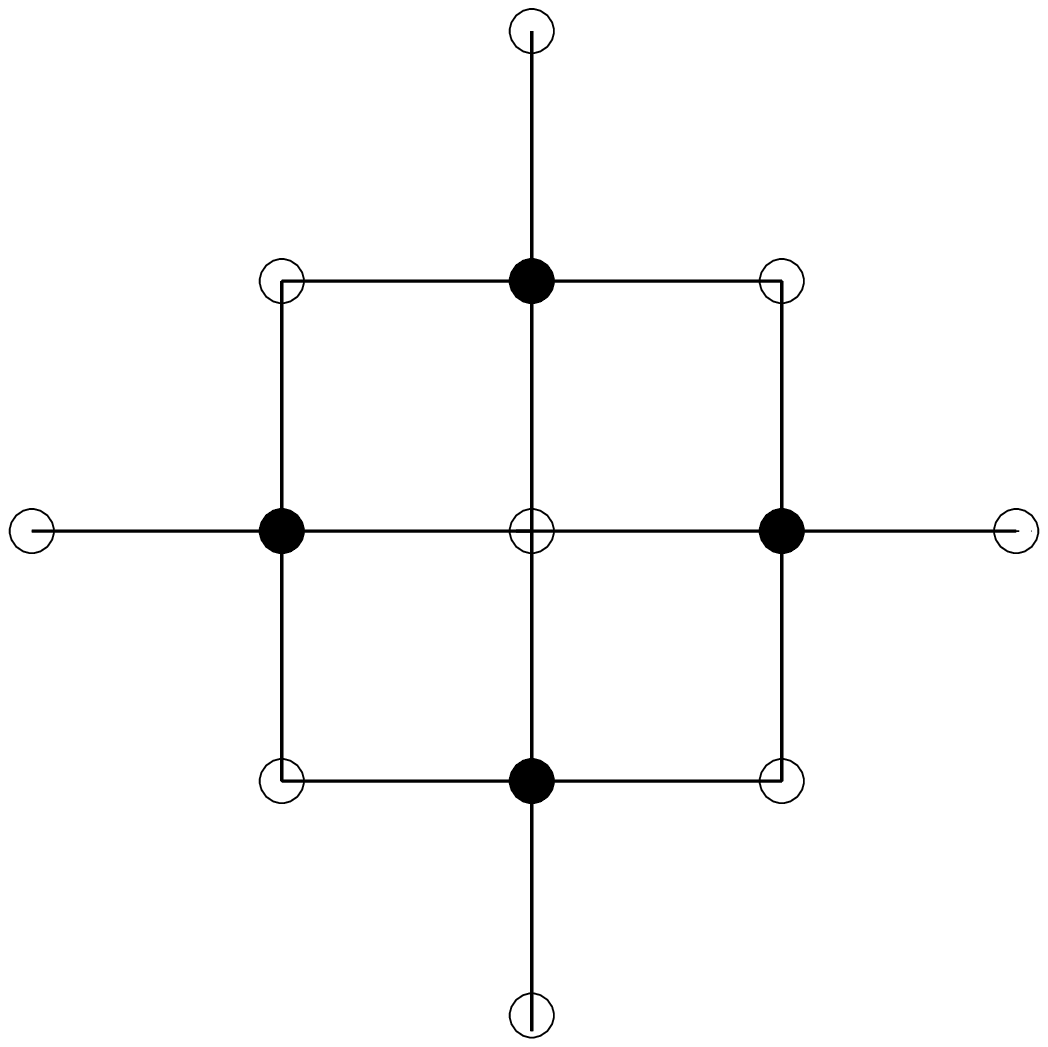} \\
contribution & $4P+\mathcal O (P^2)$ & $2P+\mathcal O (P^2)$ & $P+\mathcal O (P^2)$ & $P+\mathcal O (P^2)$ \\
multiplicity & 4 & 4 & 4 & 1 \\
\hline
subgraph & & & & \\
 & & \includegraphics[width=3.5cm]{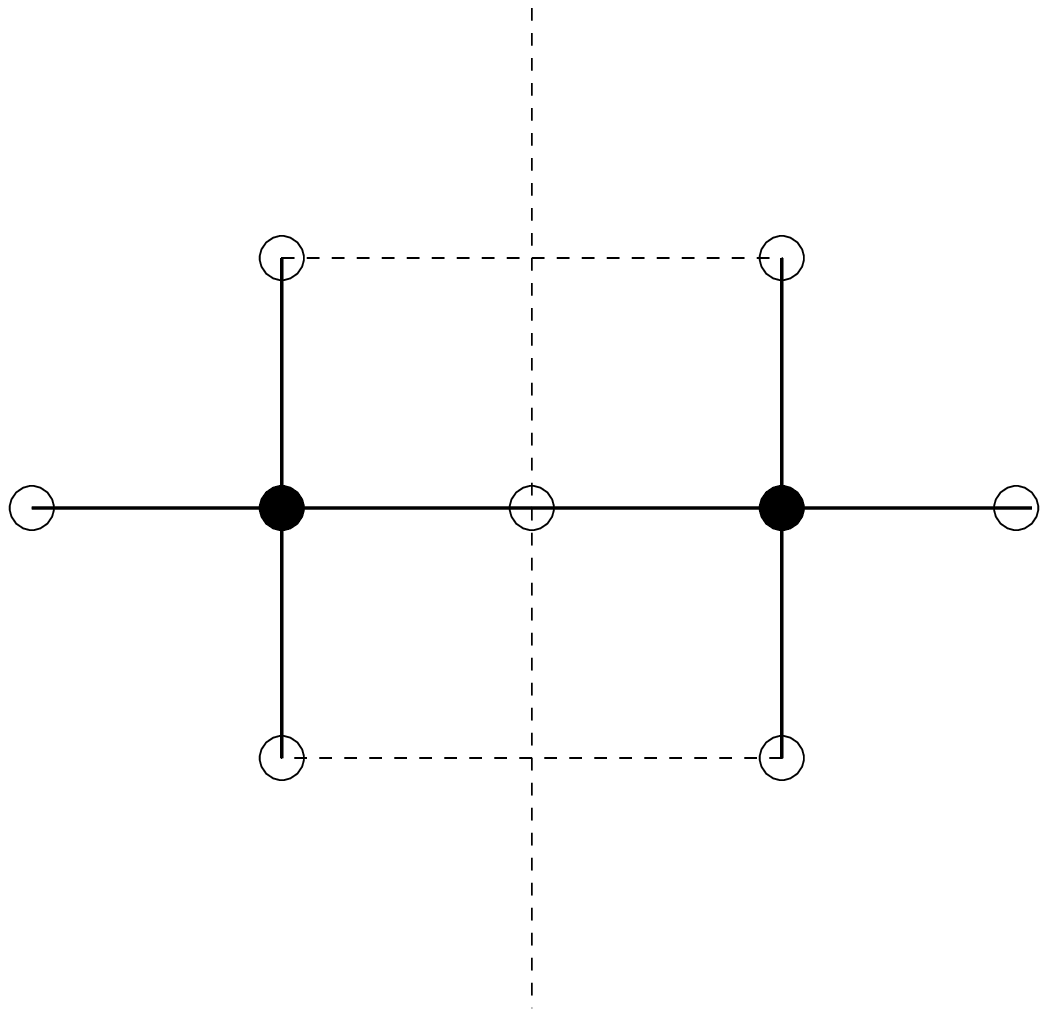} & \includegraphics[width=3.5cm]{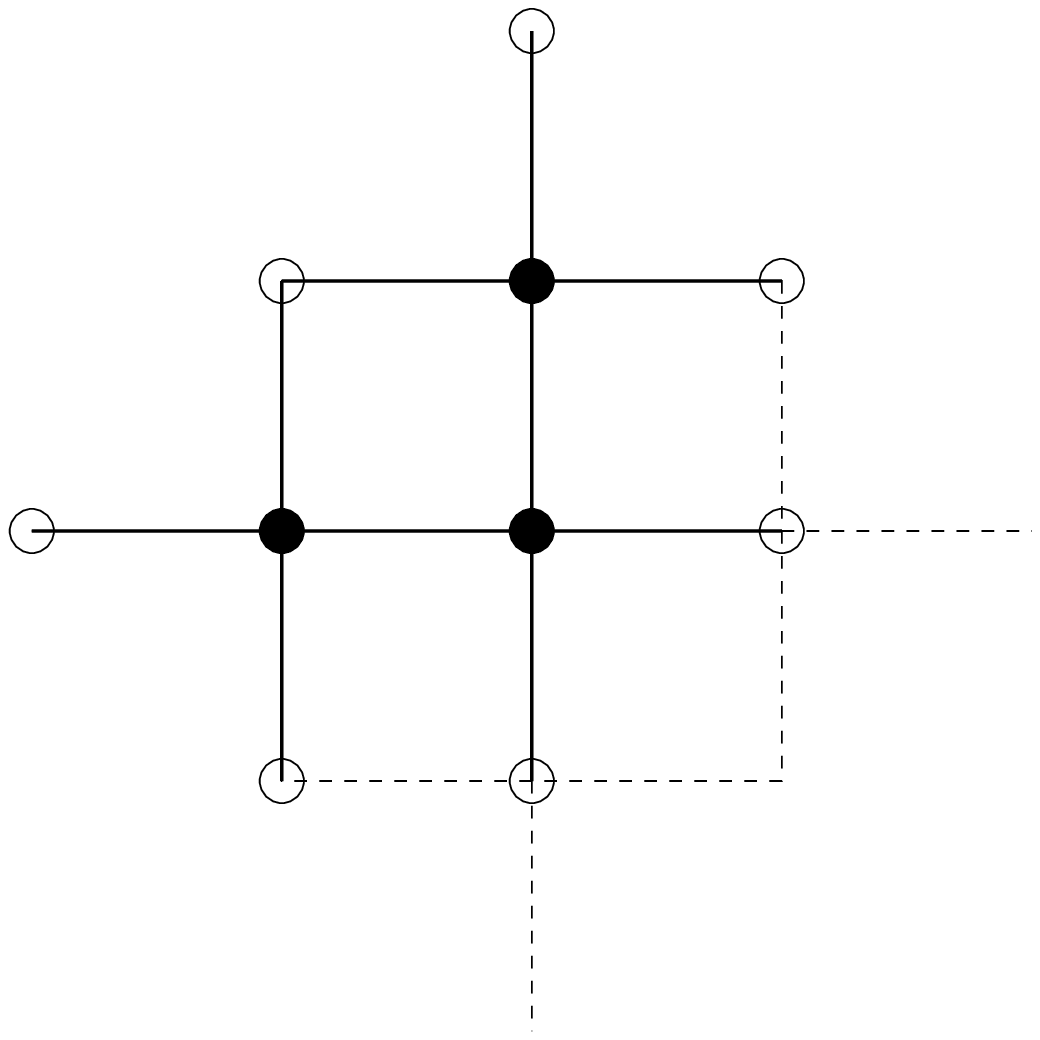} & \includegraphics[width=3.5cm]{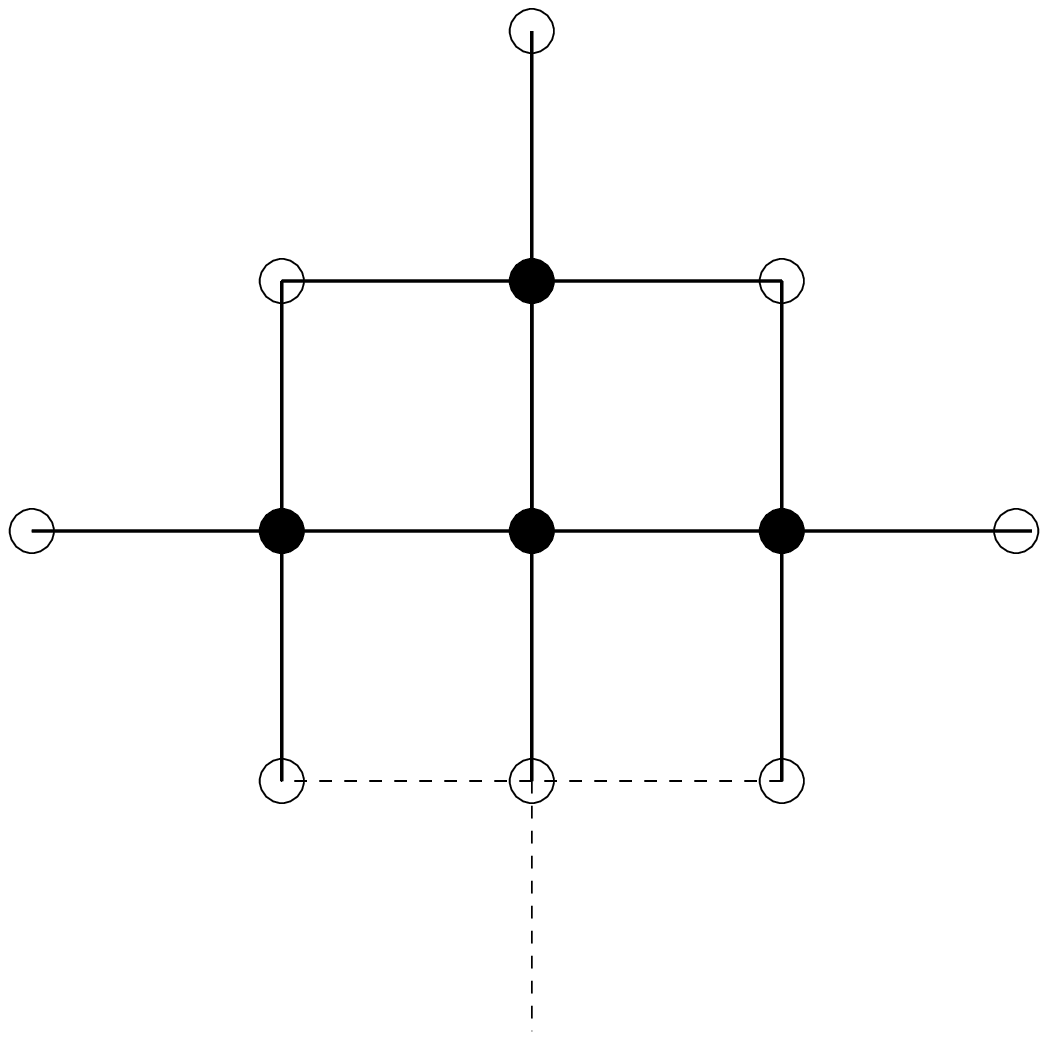} \\
contribution & & $P+\mathcal O(P^2)$ & $2P^2+\mathcal O(P^3)$ & $5P^3+\mathcal O(P^4)$ \\
multiplicity & & 2 & 4 & 4 \\
\hline
subgraph & & & & \\
 & & \includegraphics[width=3.5cm]{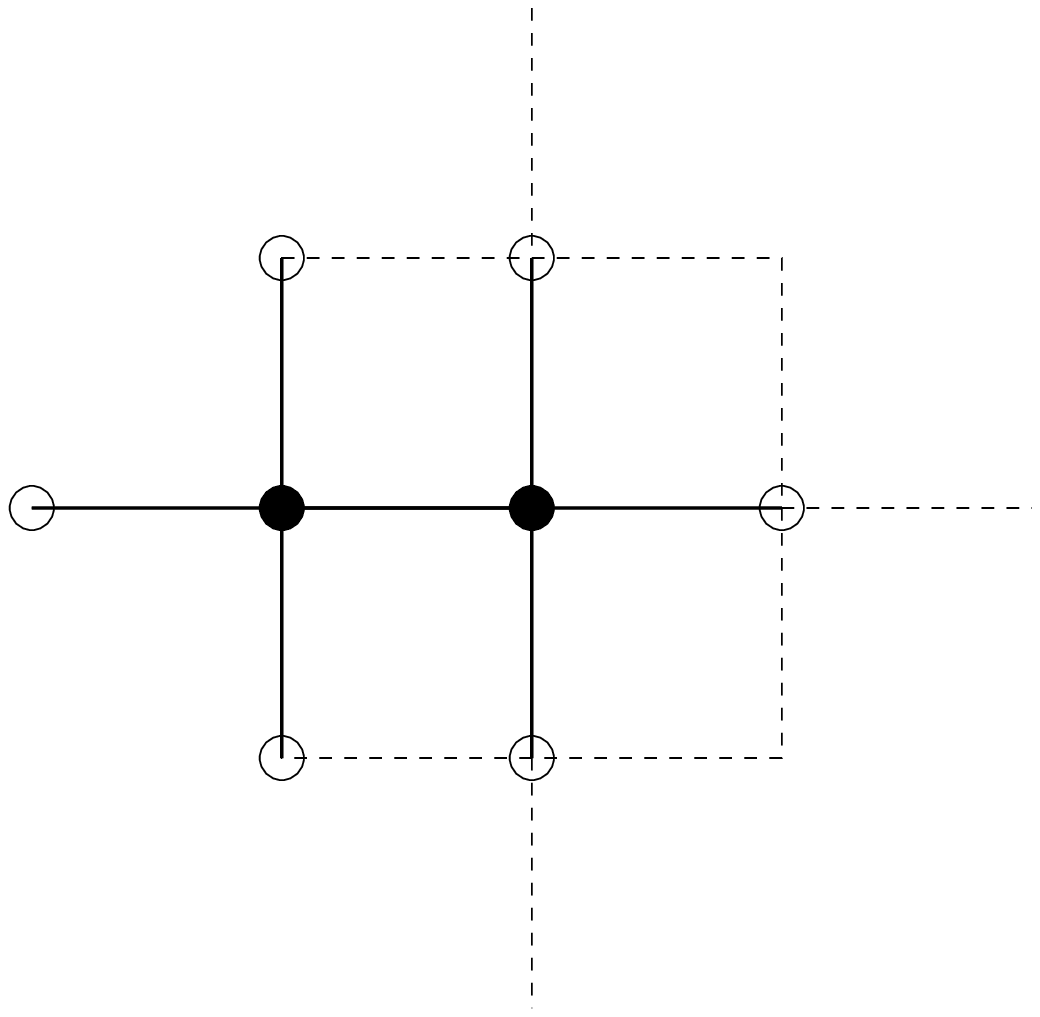} & \includegraphics[width=3.5cm]{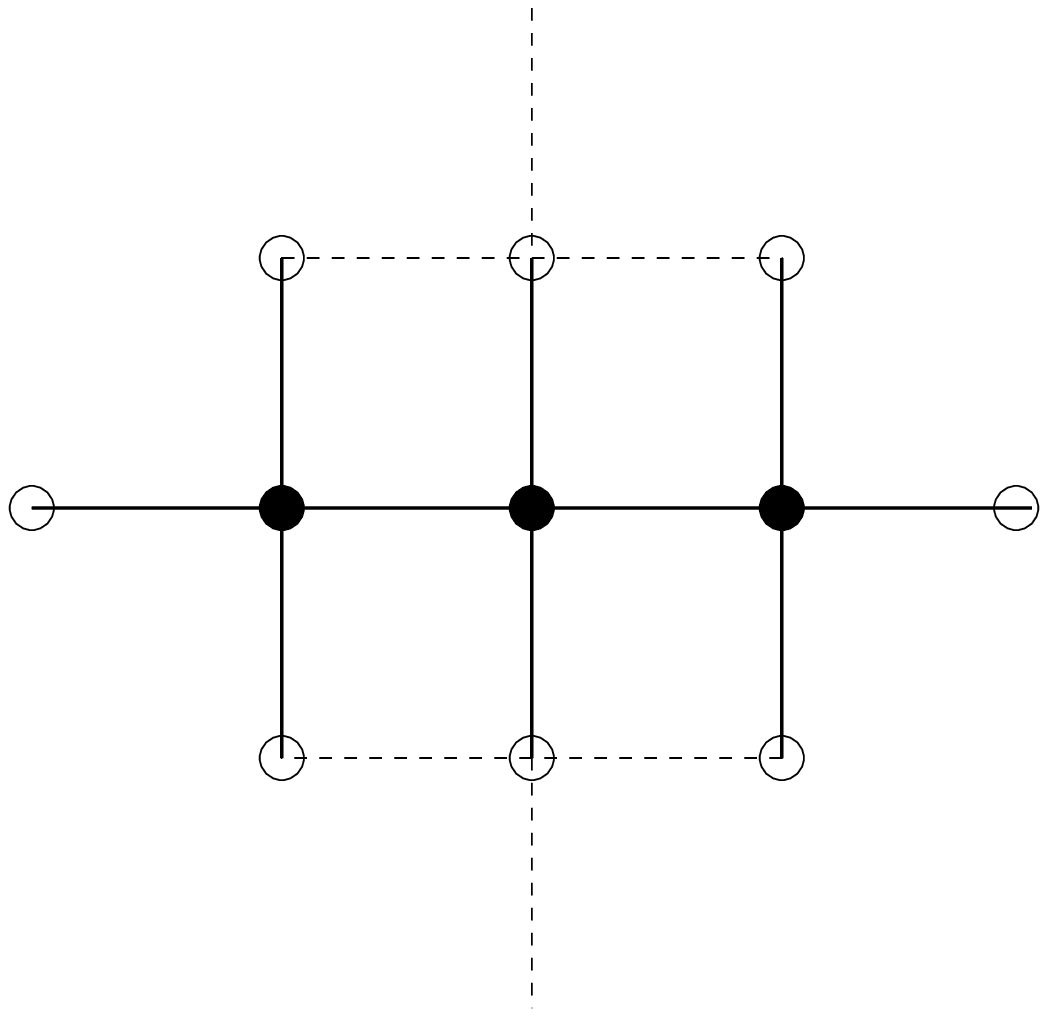} & \\
contribution & & $9P^2+\mathcal O(P^3)$ & $18P^3+\mathcal O(P^4)$ & \\
multiplicity & & 4 & 2 & \\
\hline
$\lambda\cdot\lambda^n$-coeff. & 16 & -10 & 4 & -1 \\
\end{tabular}
\end{ruledtabular}
\label{tab:sq_sub}
\end{table*}

\begin{table*}[h]
\caption{The subgraphs of all the orders for the Kagom\'e lattice. See Tab.\ \ref{tab:sq_sub} for how the coefficients are obtained and as far as further details are concerned.}
\begin{ruledtabular}
\begin{tabular}{|c||c|c|c|c|}
order ($n$) & 1 & 2 & 3 & 4 \\
\hline
subgraph & & & & \\
 & \includegraphics[width=3.5cm]{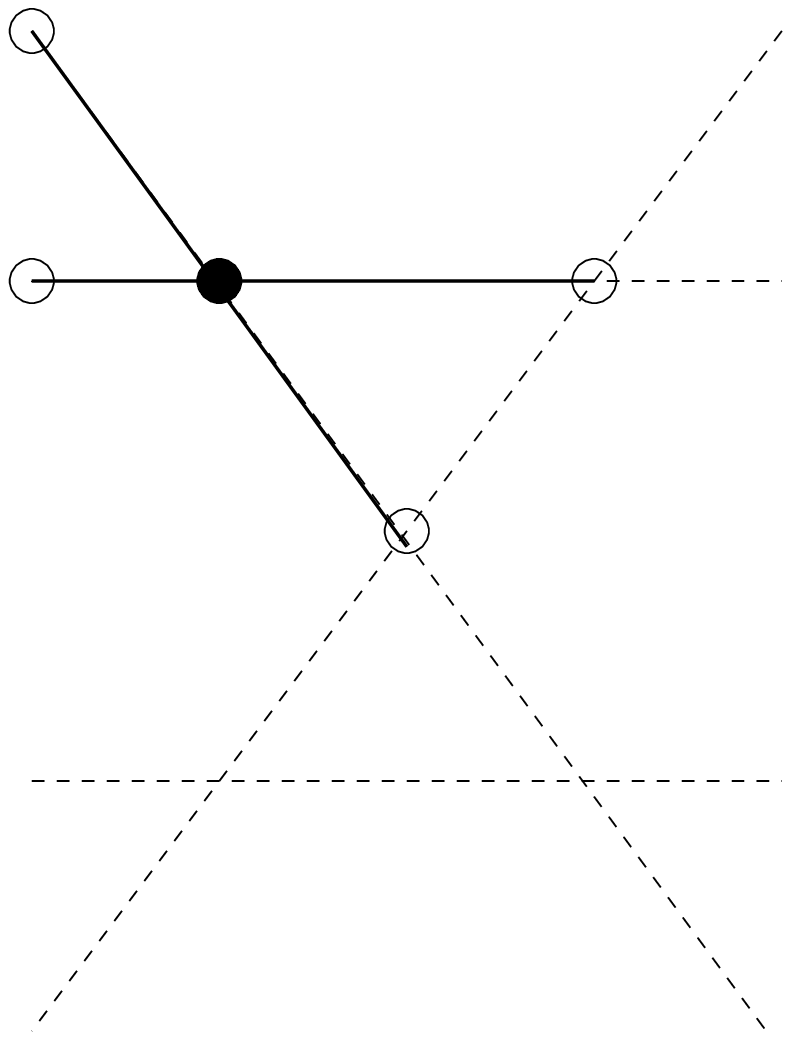} & \includegraphics[width=3.5cm]{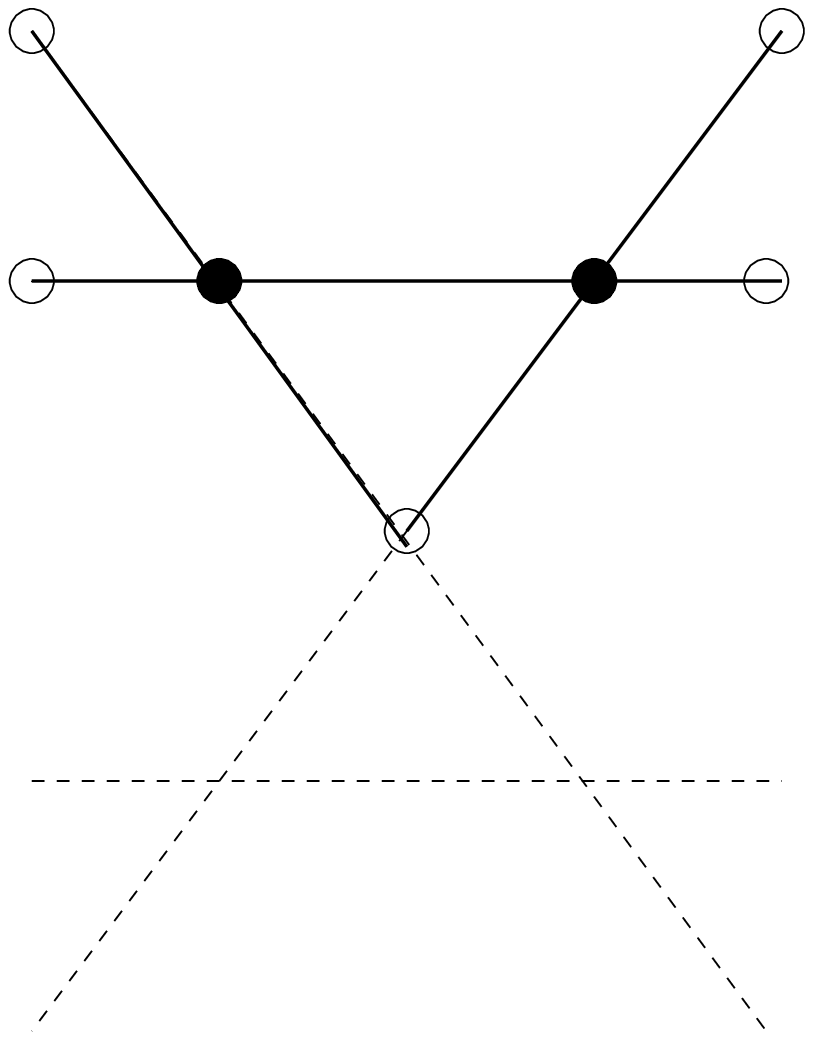} & \includegraphics[width=3.5cm]{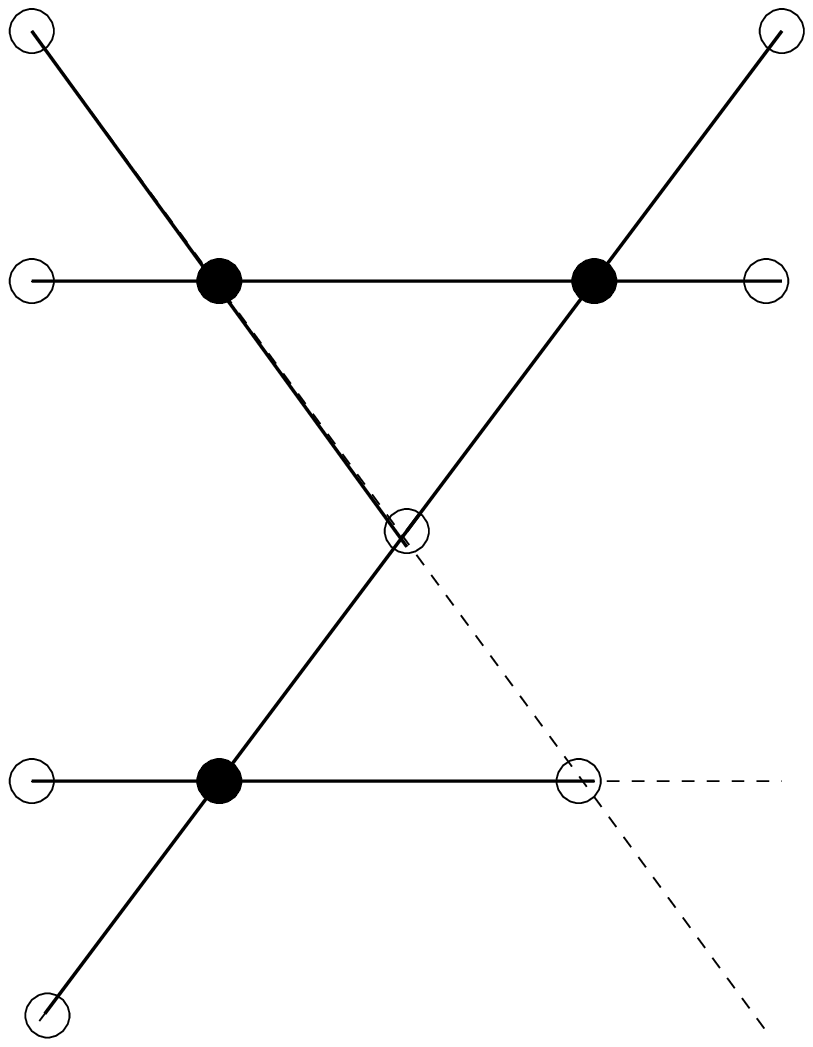} & \includegraphics[width=3.5cm]{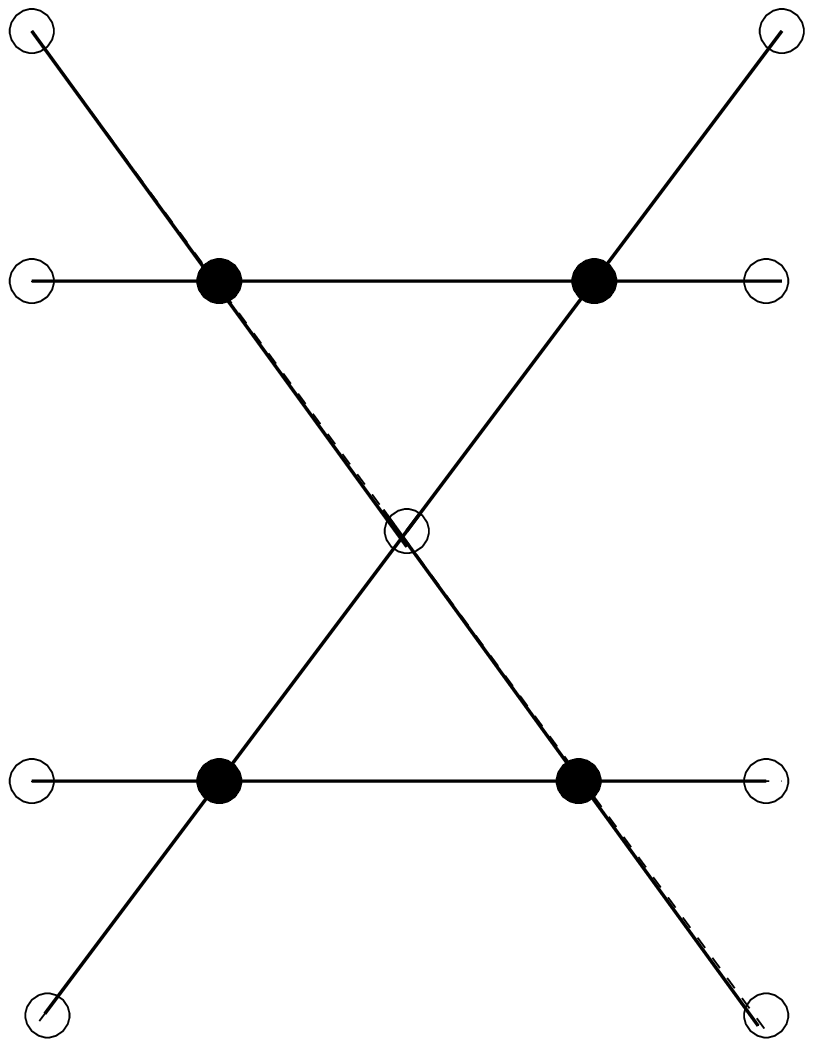} \\
contribution & $4P+\mathcal O (P^2)$ & $P+\mathcal O (P^2)$ & $P+\mathcal O (P^2)$ & $P+\mathcal O (P^2)$ \\
multiplicity & 4 & 6 & 4 & 1 \\
\hline
subgraph & & & & \\
 & & \includegraphics[width=3.5cm]{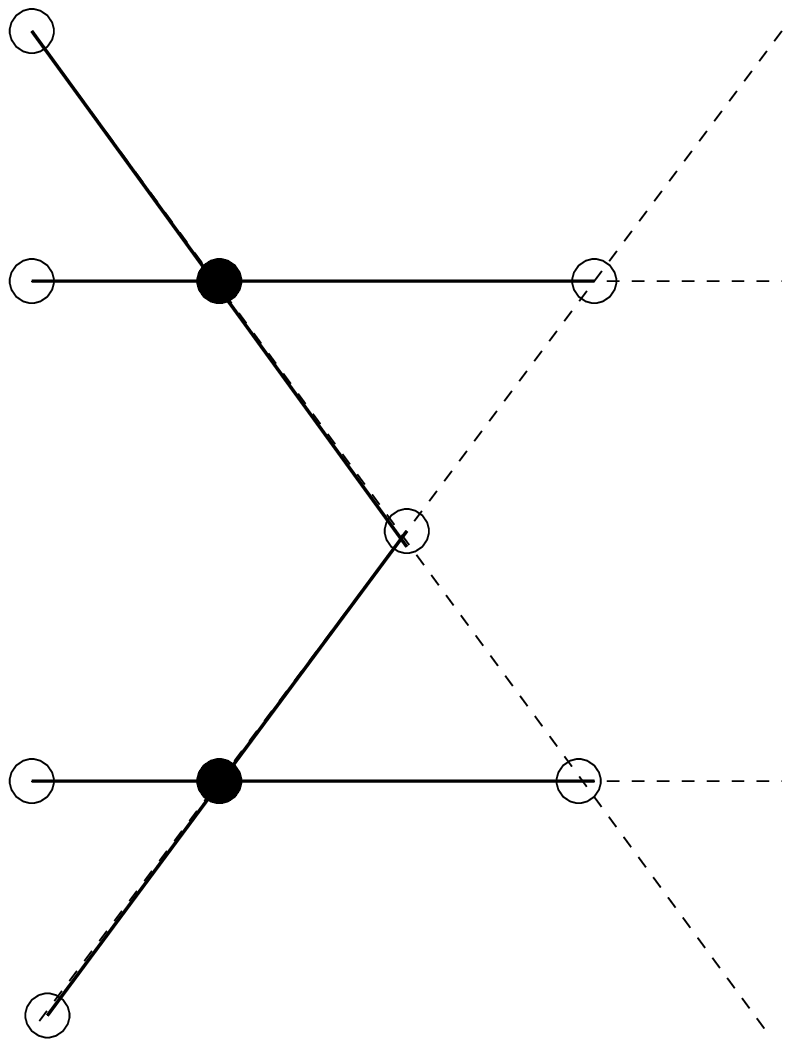} & \includegraphics[width=3.5cm]{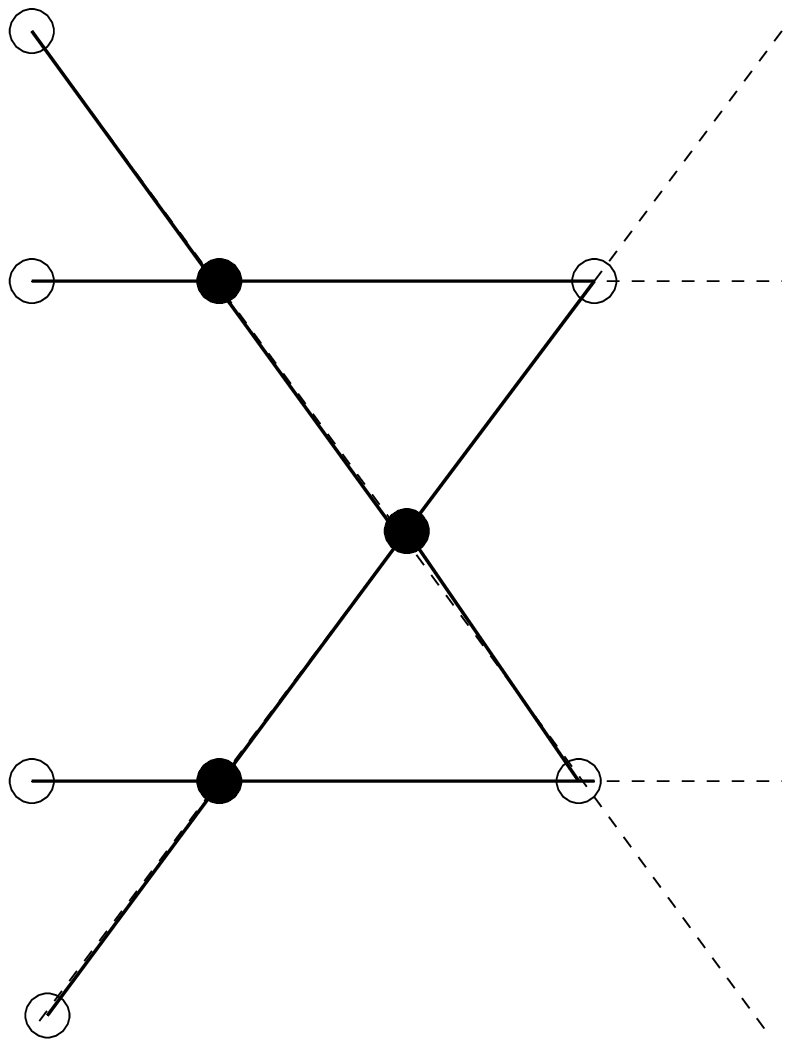} & \includegraphics[width=3.5cm]{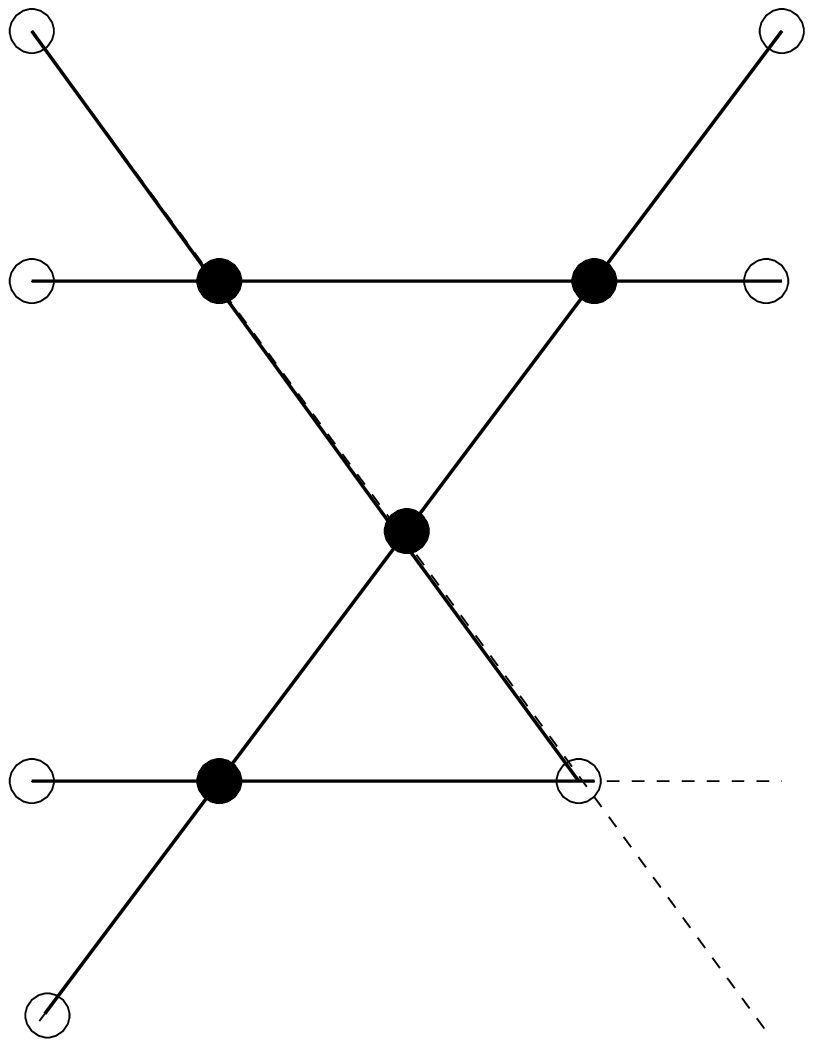} \\
contribution & & $P+\mathcal O(P^2)$ & $5P^2+\mathcal O(P^3)$ & $8P^4+\mathcal O(P^5)$ \\
multiplicity & & 4 & 4 & 4 \\
\hline
subgraph & & & & \\
 & & & \includegraphics[width=3.5cm]{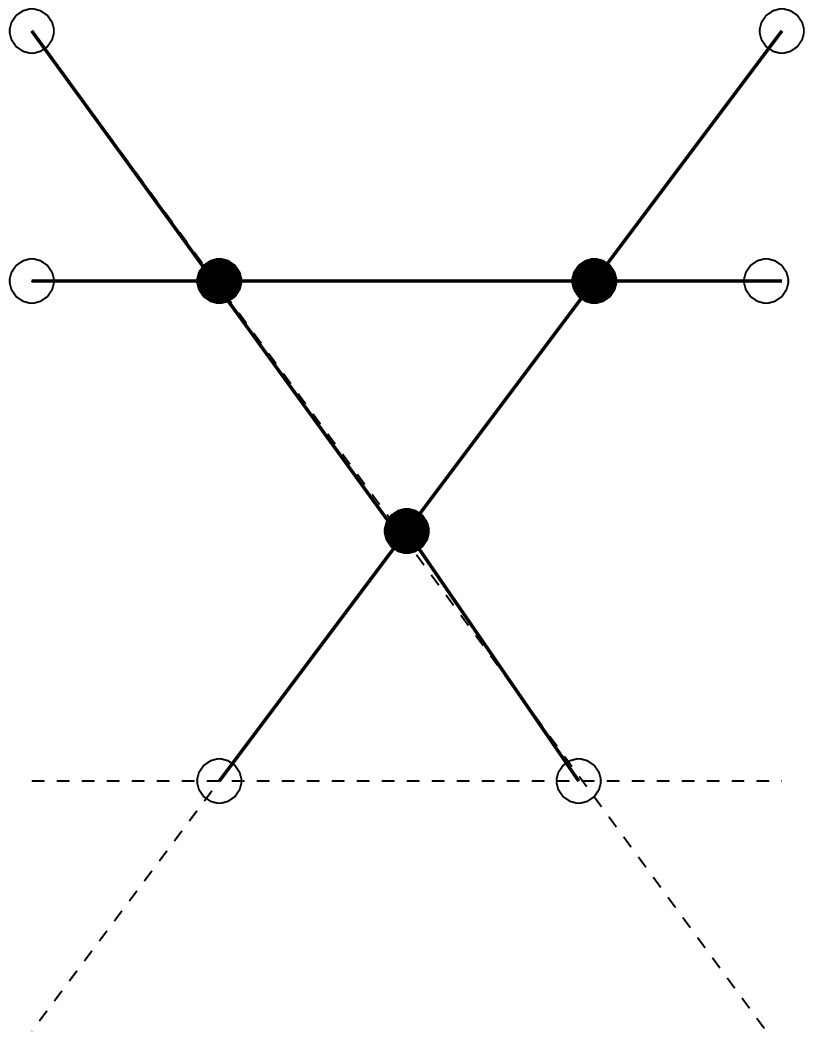} &  \\
contribution & & & $8P^3+\mathcal O(P^4)$ & \\
multiplicity & & & 2 & \\
\hline
$\lambda\cdot\lambda^n$-coeff. & 16 & -10 & 4 & -1 \\
\end{tabular}
\end{ruledtabular}
\label{tab:kag_sub}
\end{table*}

\begin{table*}[h]
\caption{The full subgraph development for the ring-type network. See Tab.\ \ref{tab:sq_sub} for the derivation of the $\lambda\cdot\lambda^n$-coefficients. With respect to the two lattices treated above, the $\lambda\cdot\lambda^2$- and $\lambda\cdot\lambda^3$-coefficients are -16 and 6.}
\begin{ruledtabular}
\begin{tabular}{|c||c|c|c|c|}
order ($n$) & 1 & 2 & 3 & 4 \\
\hline
subgraph & & & & \\
 & \includegraphics[width=3.5cm]{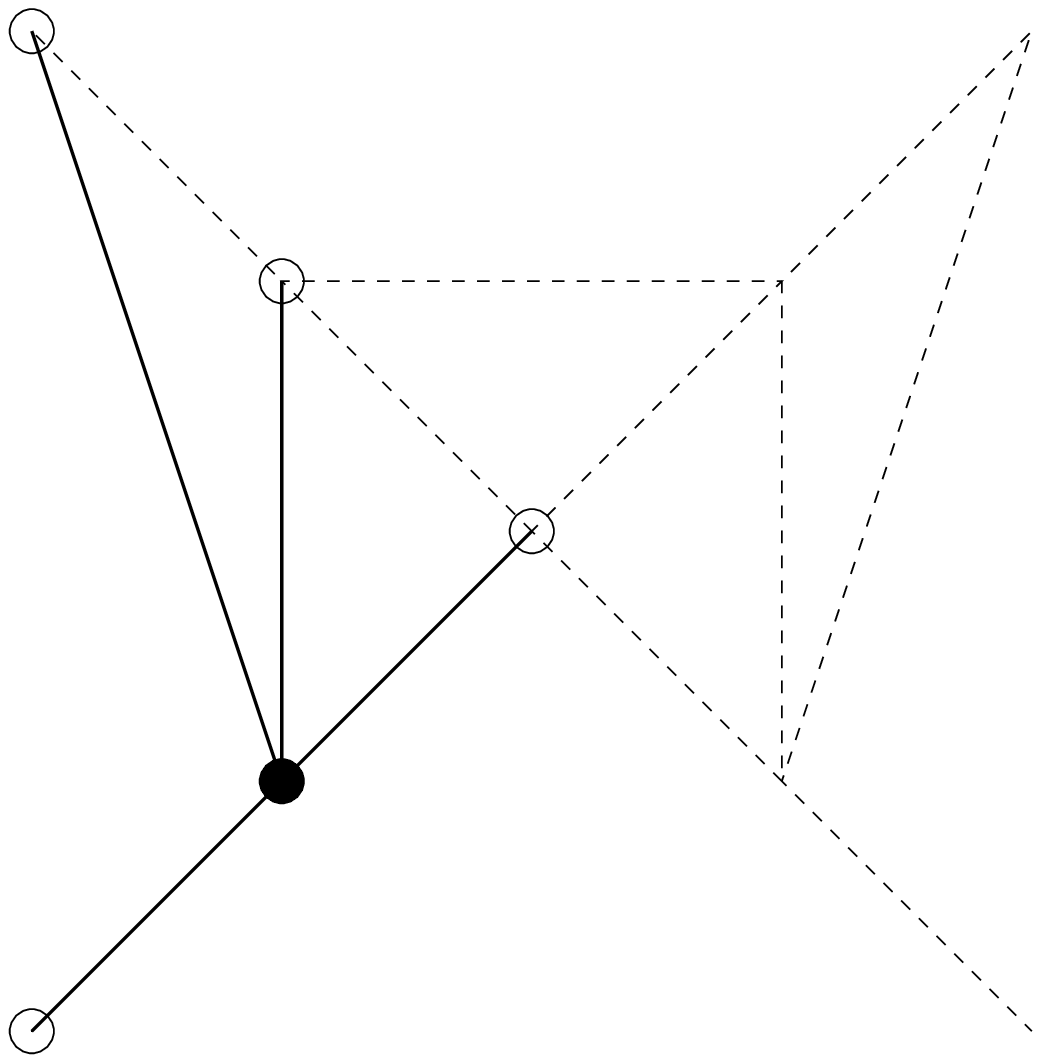} & \includegraphics[width=3.5cm]{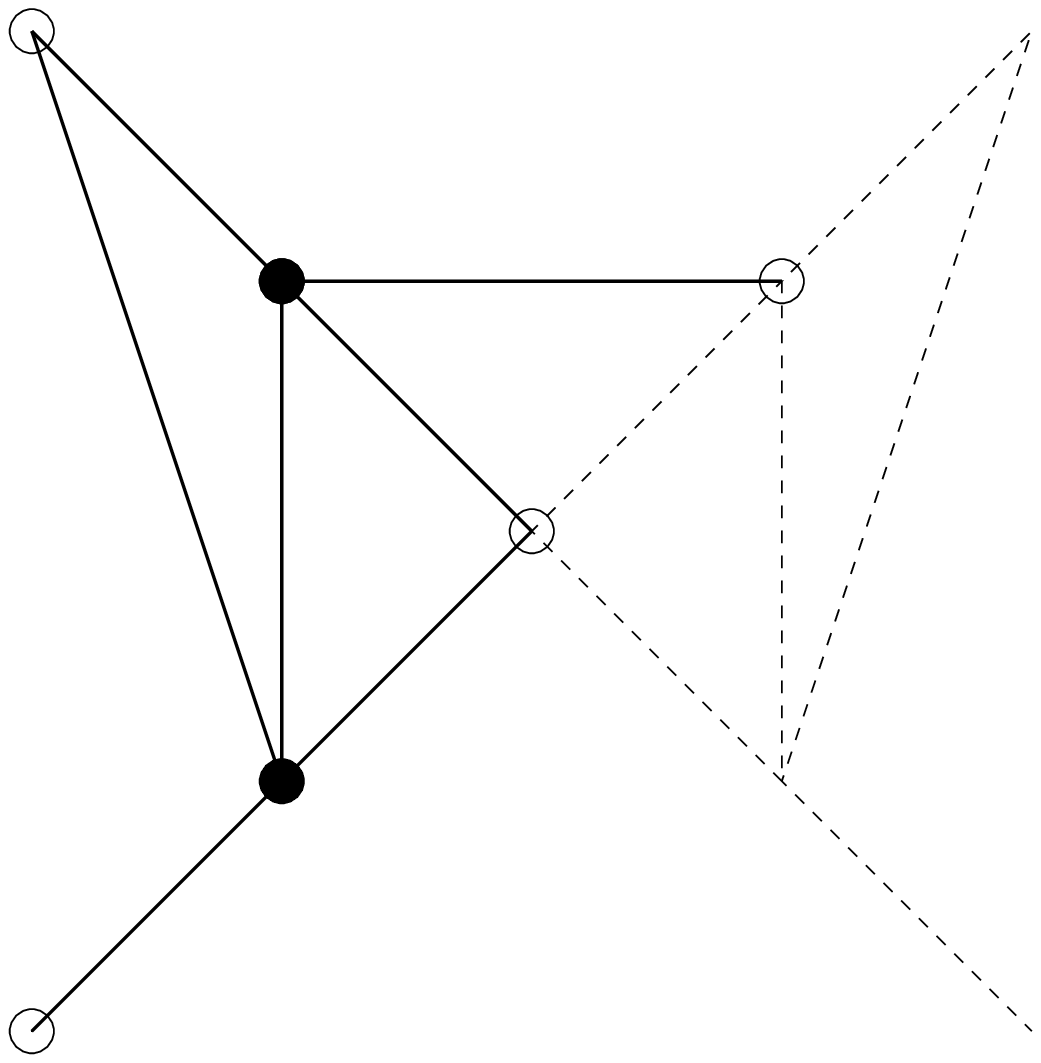} & \includegraphics[width=3.5cm]{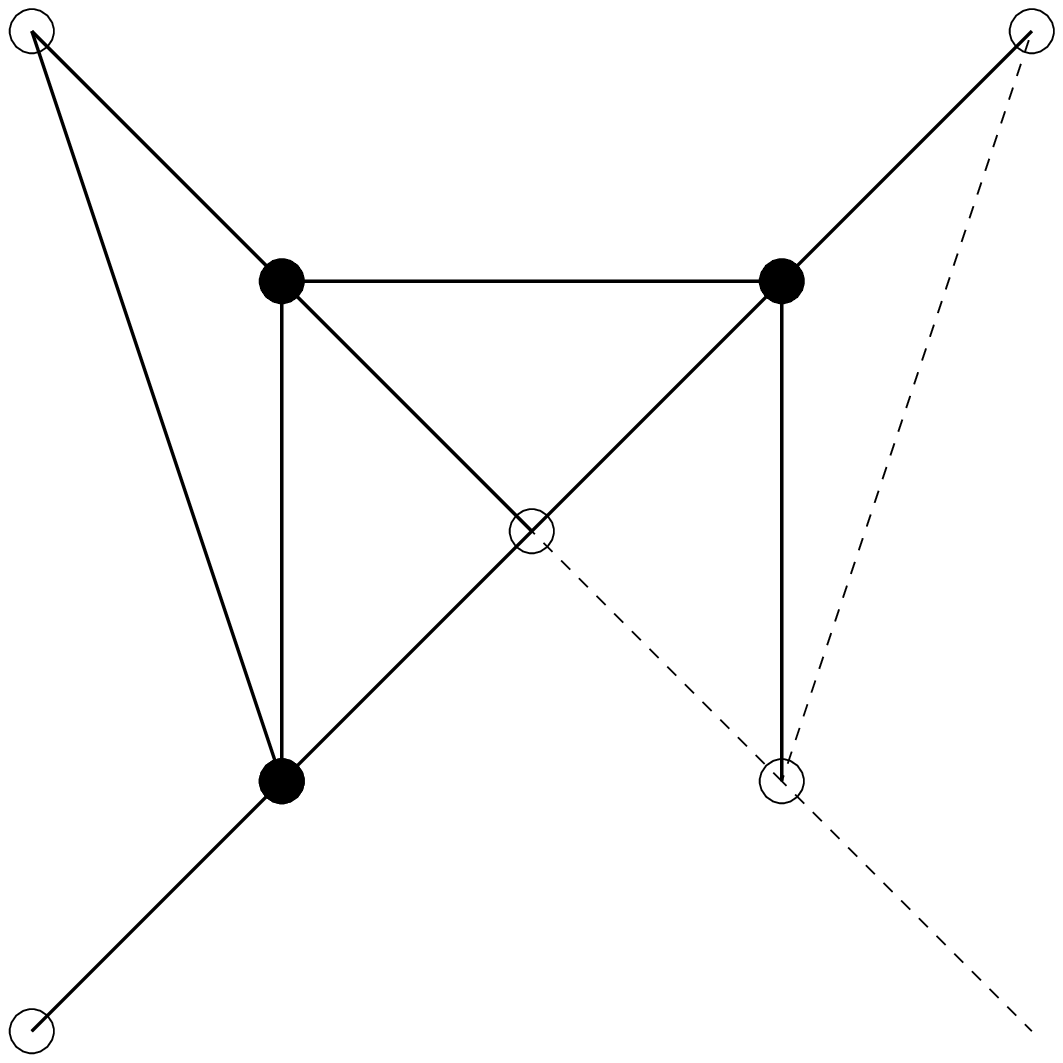} & \includegraphics[width=3.5cm]{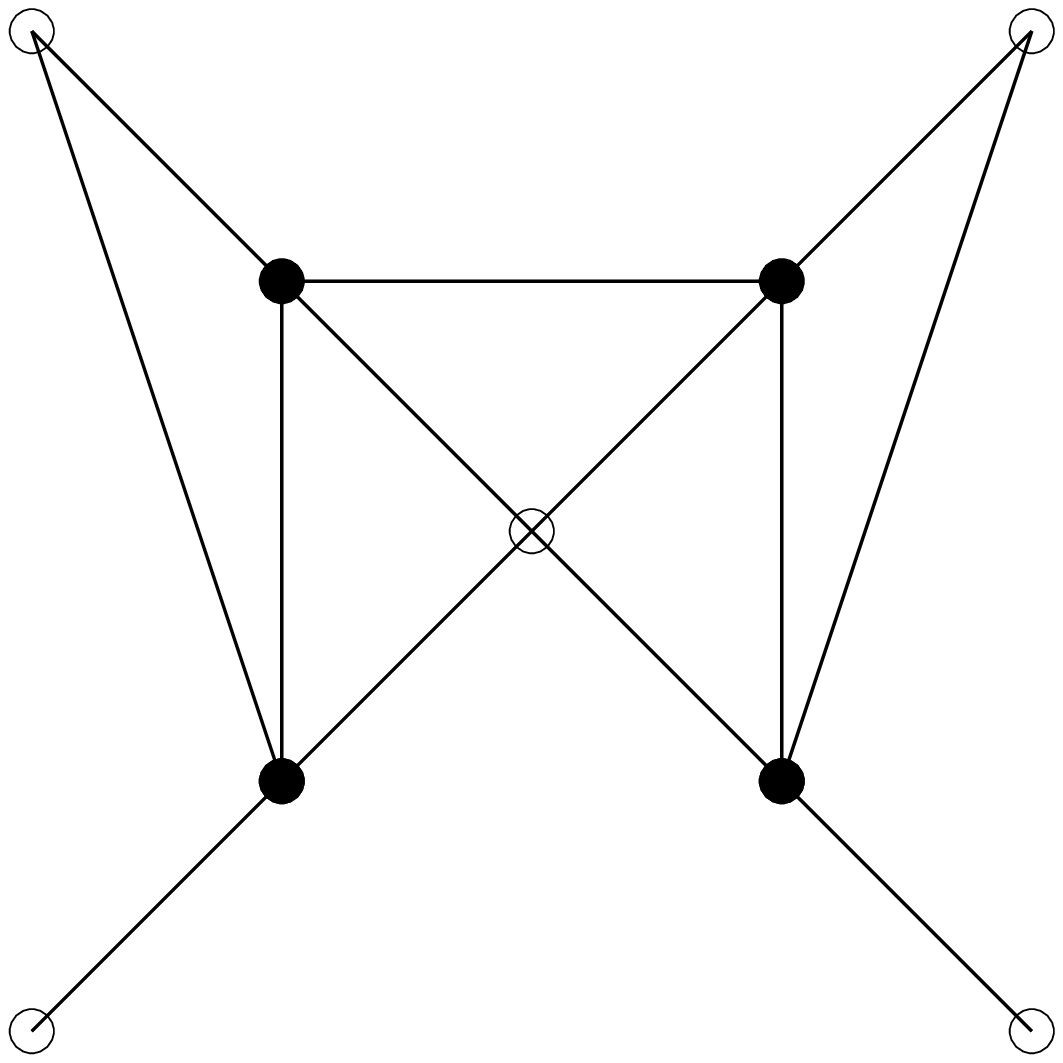} \\
contribution & $4P+\mathcal O (P^2)$ & $2P+\mathcal O (P^2)$ & $P+\mathcal O (P^2)$ & $P+\mathcal O (P^2)$ \\
multiplicity & 4 & 4 & 4 & 1 \\
\hline
subgraph & & & & \\
 & & \includegraphics[width=3.5cm]{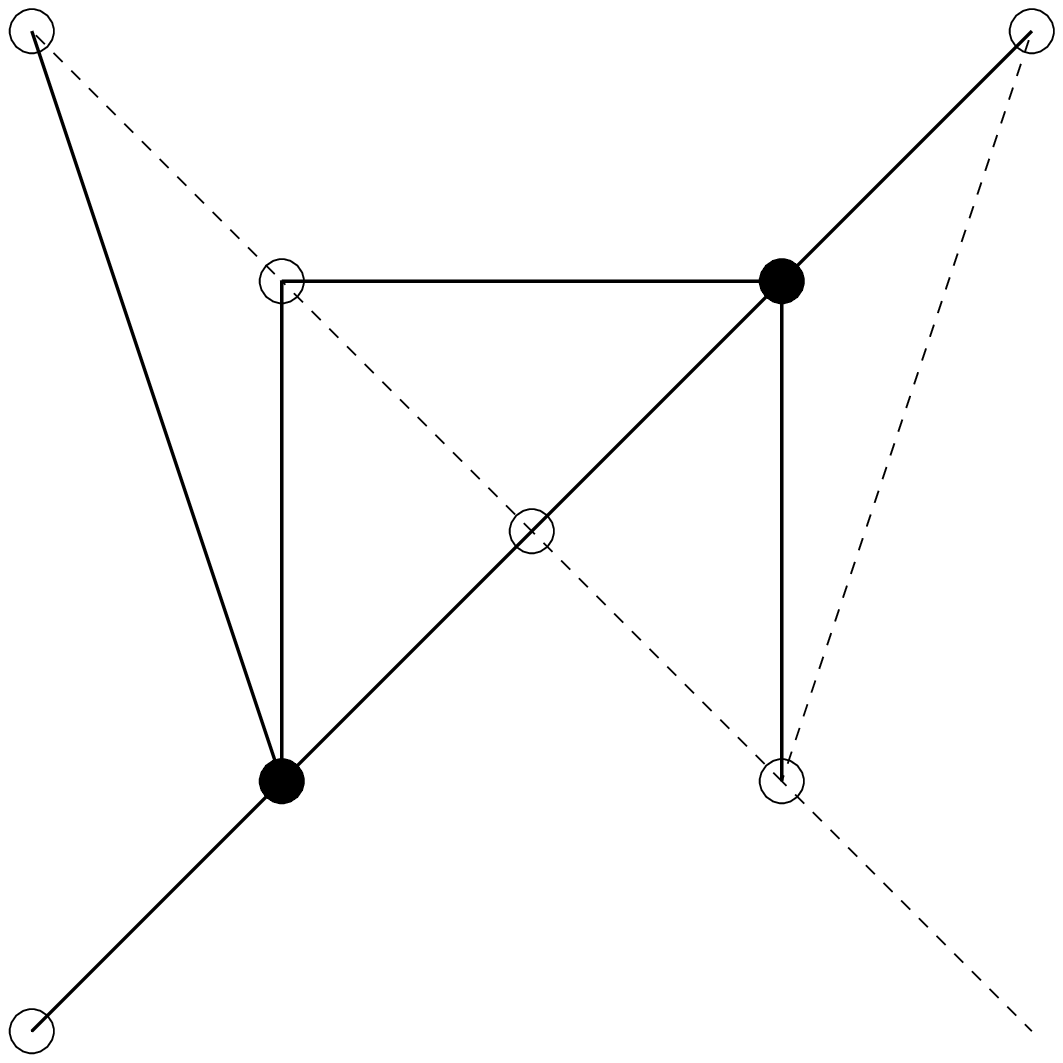} & \includegraphics[width=3.5cm]{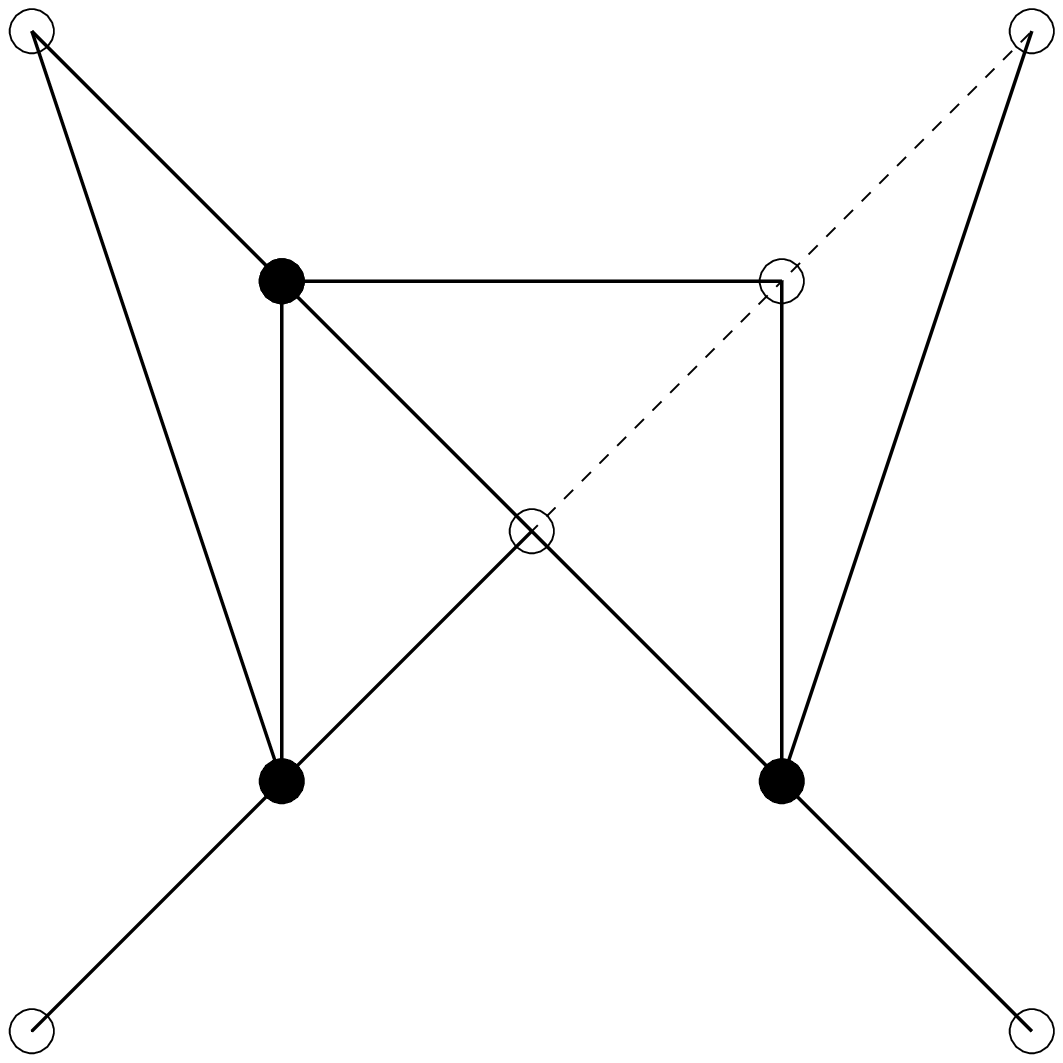} & \includegraphics[width=3.5cm]{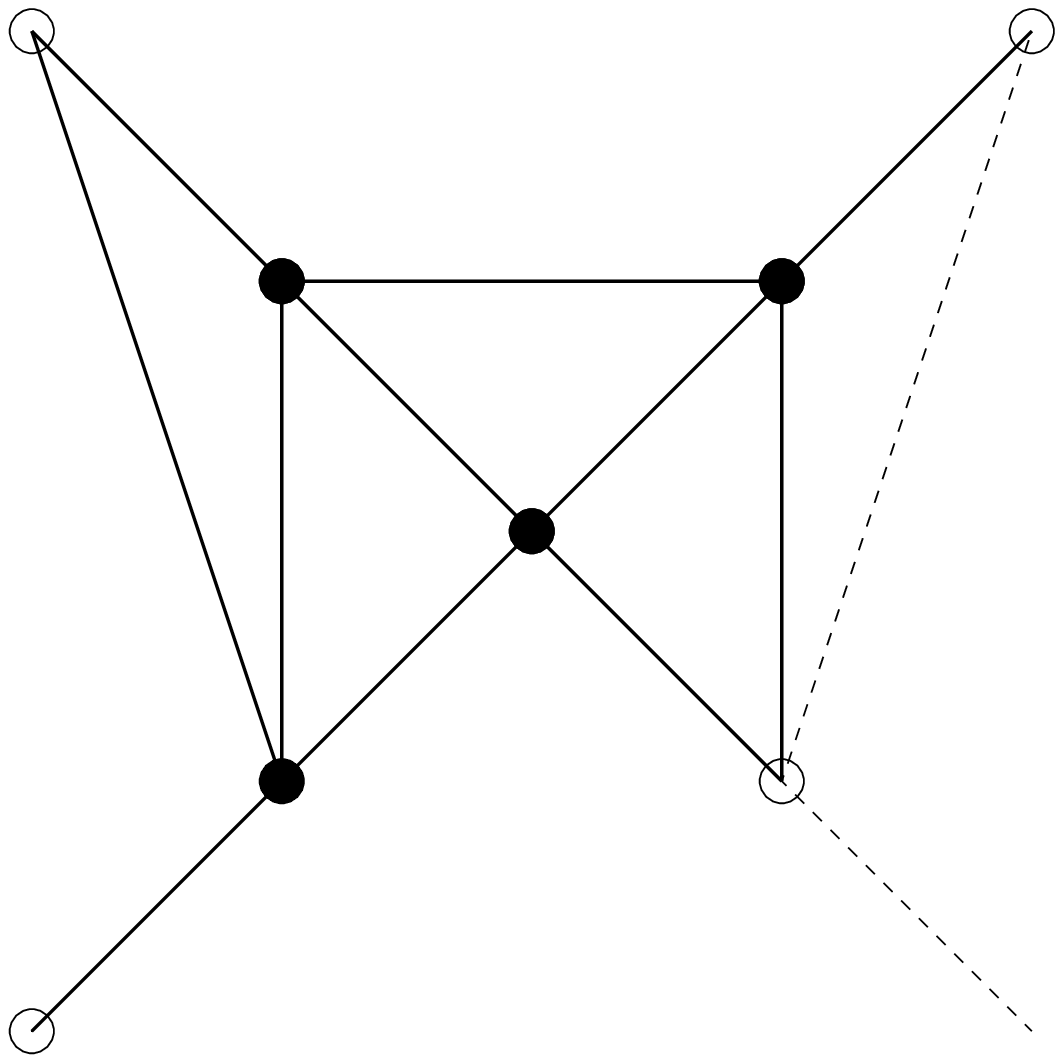} \\
contribution & & $2P+\mathcal O(P^2)$ & $P+\mathcal O(P^2)$ & $P^2+\mathcal O(P^3)$ \\
multiplicity & & 2 & 2 & 2 \\
\hline
subgraph & & & & \\
 & & \includegraphics[width=3.5cm]{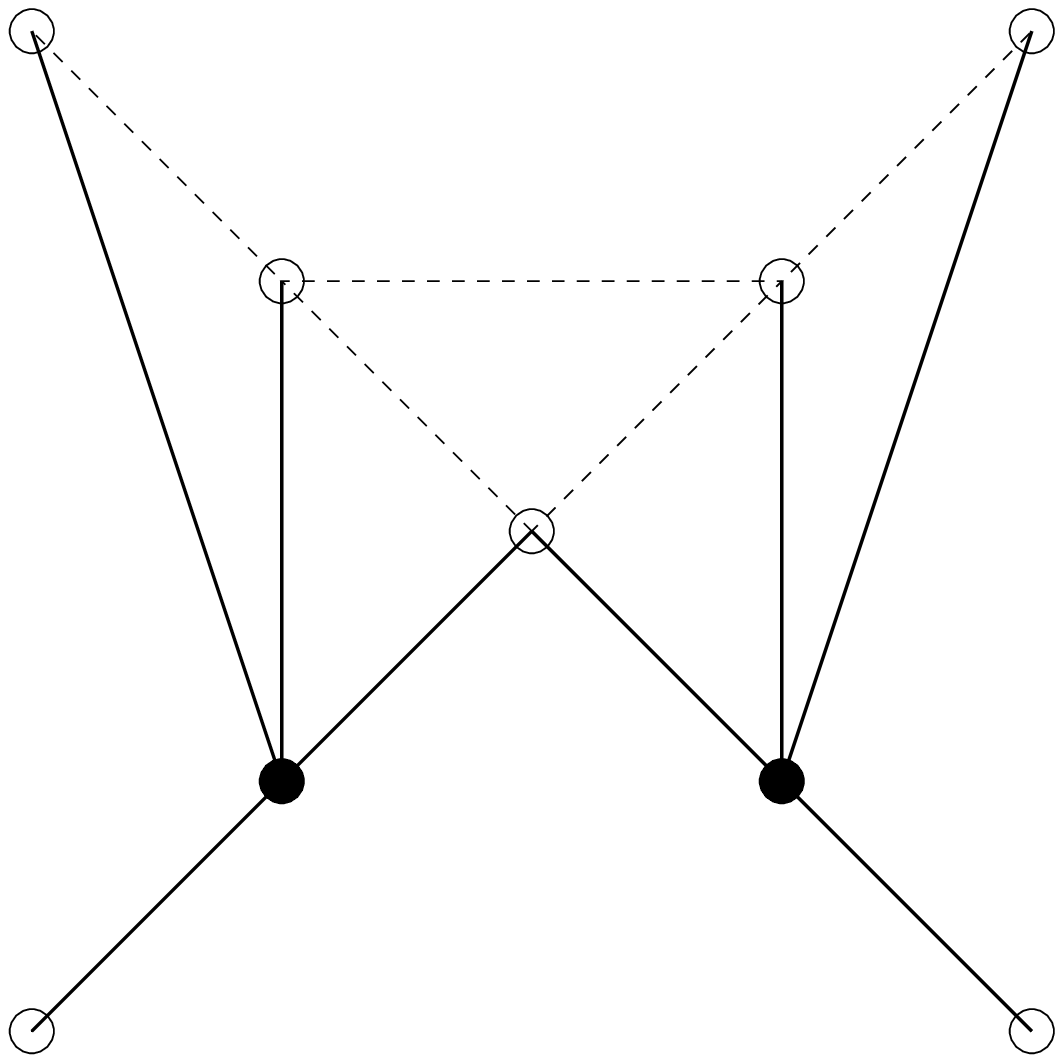} & \includegraphics[width=3.5cm]{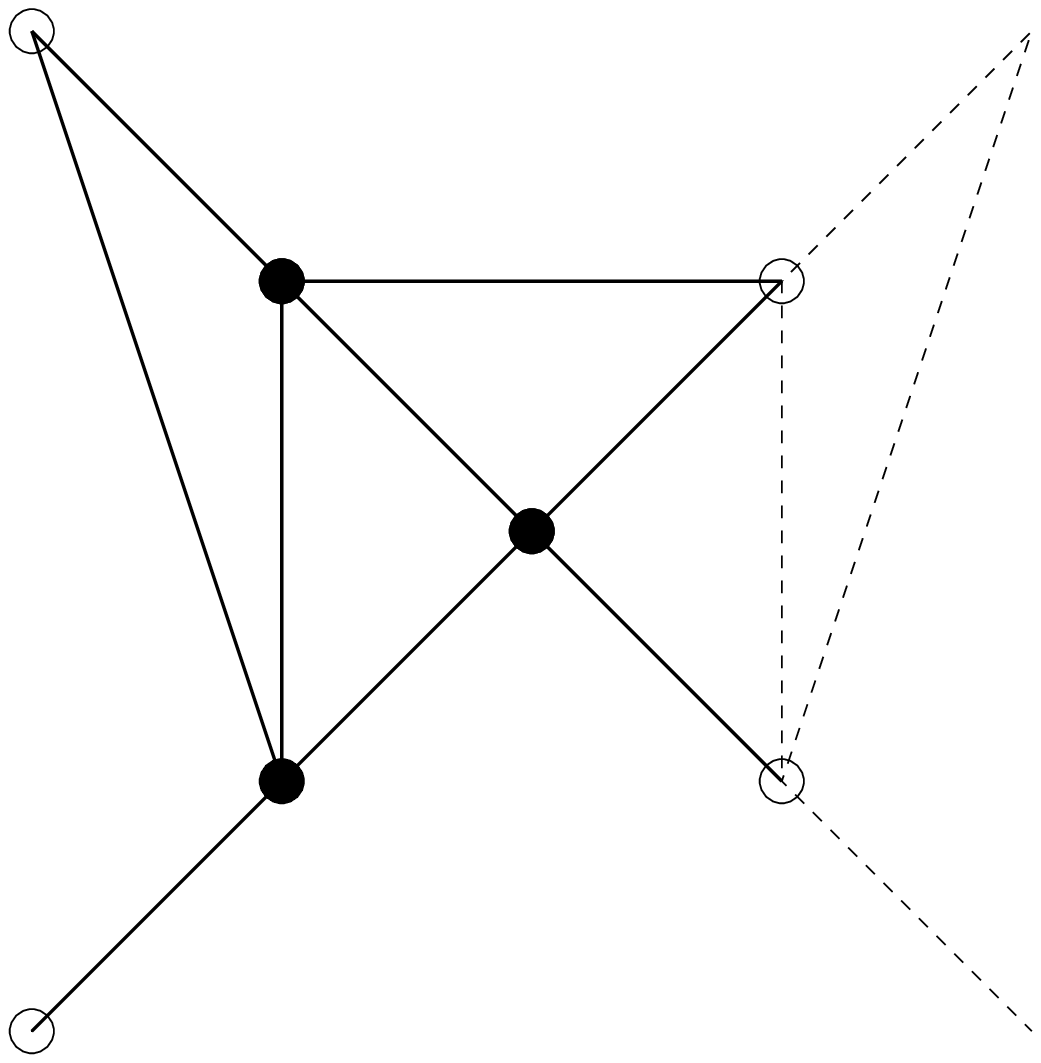} & \includegraphics[width=3.5cm]{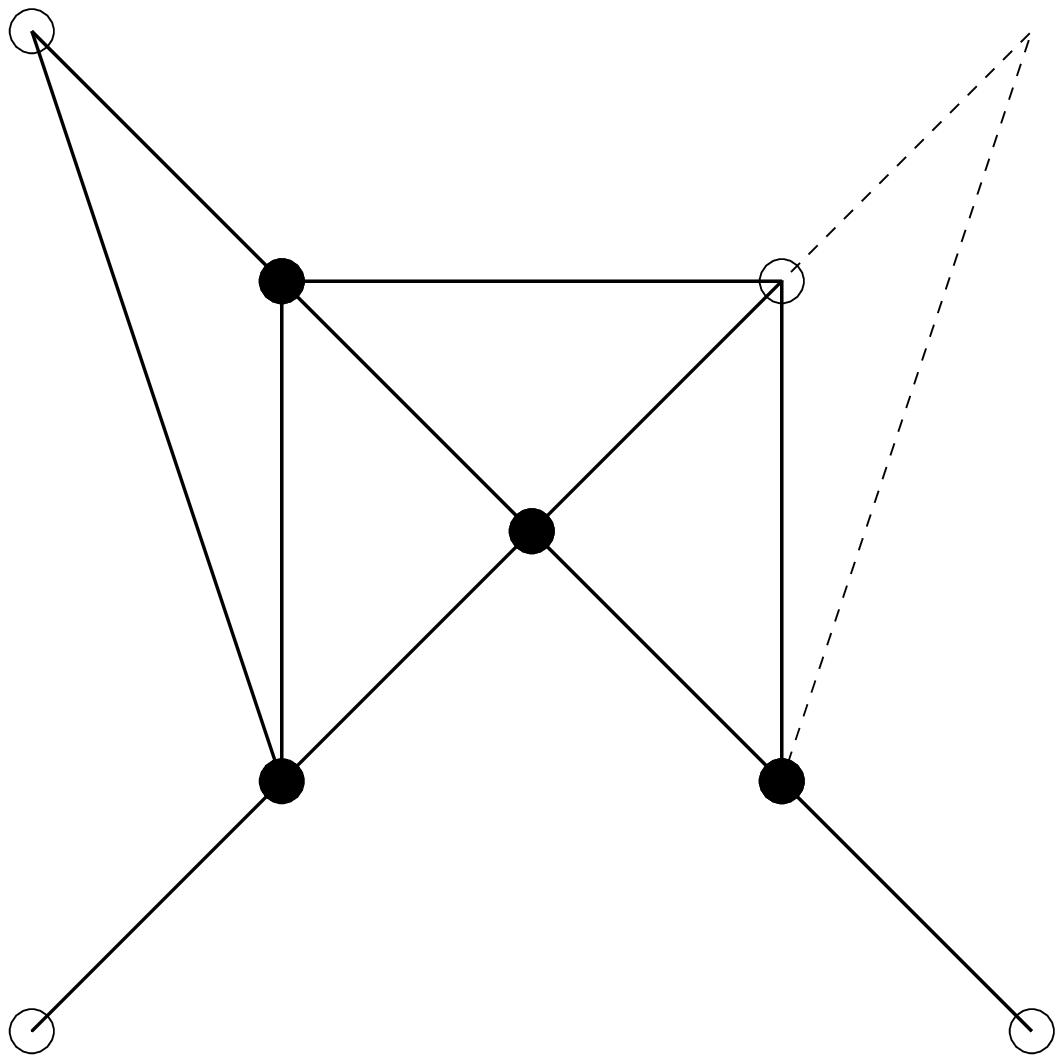} \\
contribution & & $P+\mathcal O(P^2)$ & $3P^2+\mathcal O(P^3)$ & $2P^2+\mathcal O(P^3)$ \\
multiplicity & & 1 & 3 & 2 \\
\hline
subgraph & & & & \\
 & & \includegraphics[width=3.5cm]{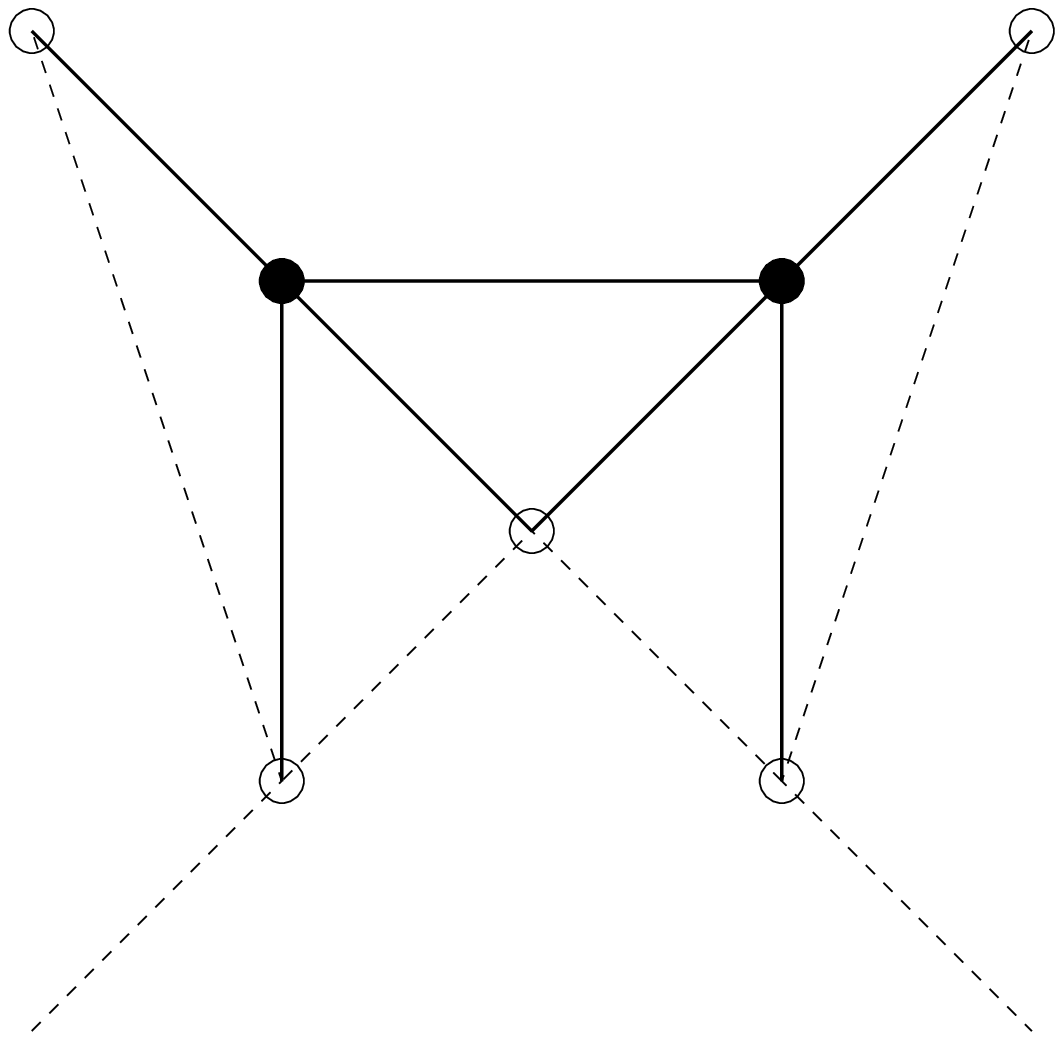} & \includegraphics[width=3.5cm]{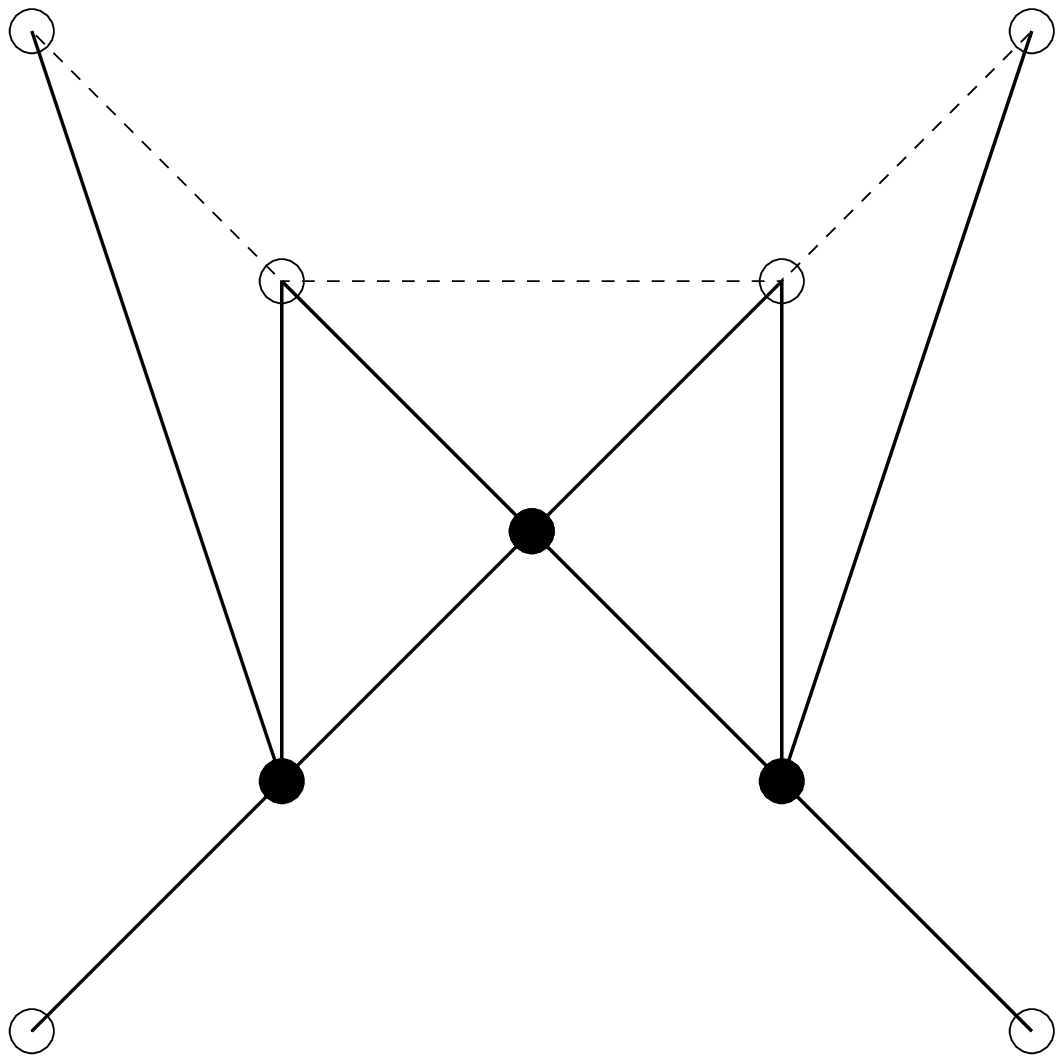} &  \\
contribution & & $P+\mathcal O(P^2)$ & $5P^2+\mathcal O(P^3)$ & \\
multiplicity & & 3 & 1 & \\
\hline
$\lambda\cdot\lambda^n$-coeff. & 16 & -16 & 6 & -1 \\
\end{tabular}
\end{ruledtabular}
\label{tab:ring_sub}
\end{table*}


\begin{thebibliography}{99}

%{\bf REFERENCES}

%\vspace{4mm}

\bibitem{milgram}
S. Milgram, Psychology Today {\bf 1}, 60 (1967).

\bibitem{erdoes}
P. Erd\"os and A. R\'enyi, Publ. Math. {\bf 6}, 290 (1959).

\bibitem{watts}
D.J. Watts and S.H. Strogatz, Nature (London) {\bf 393}, 440 (1998).

\bibitem{pastor2}
R. Pastor-Satorras, A. V\'azquez and A. Vespignani, Phys. Rev. Lett. {\bf 87}, 258701 (2001).

\bibitem{yook}
S. Yook, H. Jeong and A.-L. Barab\'asi, e-print cond-mat/0107417 (2002).

\bibitem{albert2}
R. Albert, H. Jeong and A.-L. Barab\'asi, Nature {\bf 401}, 130 (1999).

\bibitem{wagner}
A. Wagner and D. Fell, Technical Report No. 00-07-041, Santa Fe Institute (2000).

\bibitem{montoya}
J. Montoya and R. Sol\'e, e-print cond-mat/0011195 (2000).

\bibitem{barabasi}
A.-L. Barab\'asi and R. Albert, Science {\bf 286}, 509 (1999).

\bibitem{albert}
R. Albert and A.-L. Barab\'asi, Rev. Mod. Phys. {\bf 74}, 47 (2002).

\bibitem{dorogovtsev}
S.N. Dorogovtsev and J.F.F. Mendes, Adv. Phys. {\bf 51}, 1079 (2002).

\bibitem{faloutsos}
M. Faloutsos, P. Faloutsos and C. Faloutsos, Comput. Comm. Rev. {\bf 29}, 251 (1999).

\bibitem{liljeros}
F. Liljeros, C.R. Edling, L.A.N. Amaral, H.E. Stanley and Y. {\AA}berg, Nature (London) {\bf 411}, 907 (2001). 

\bibitem{pastor}
R. Pastor-Satorras and A. Vespignani, Phys. Rev. Lett. {\bf 86}, 3200 (2001). 

\bibitem{newman1}
M.E.J. Newman, Phys. Rev. Lett. {\bf 89}, 208701 (2002).

\bibitem{boguna1}
M. Bogu\~n\'a and R. Pastor-Satorras, Phys. Rev. E {\bf 66}, 047104 (2002).

\bibitem{boguna2}
M. Bogu\~n\'a, R. Pastor-Satorras and A. Vespignani, Phys. Rev. Lett. {\bf 90}, 028701 (2003).

\bibitem{caldarelli}
G. Caldarelli, R. Pastor-Satorras and A. Vespignani, e-print cond-mat/0212026 (2002).

\bibitem{vazquez}
A. V\'azquez, R. Pastor-Satorras and A. Vespignani, Phys. Rev. E {\bf 65}, 066130 (2002).

\bibitem{ravasz}
E. Ravasz and A.-L. Barab\'asi, Phys. Rev. E {\bf 67}, 026112 (2003).

\bibitem{gleiss}
P.M. Gleiss, P.F. Stadler, A. Wagner and D.A. Fell, Adv. Complex Syst. {\bf 4}, 207 (2001).

\bibitem{bianconi}
G. Bianconi and A. Capocci, Phys. Rev. Lett. {\bf 90}, 078701 (2003).

\bibitem{newman2}
M.E.J. Newman, Phys. Rev. E {\bf 68}, 026121 (2003).

\bibitem{matsuda}
H. Matsuda, N. Ogita, A. Sasaki and K. Sato, Prog. Theor. Phys. {\bf 88}, 1035 (1992). 

\bibitem{vanBaalen}
M. Van Baalen, in {\it The geometry of ecological interactions: simplifying spatial complexity} (edited by U. Dieckmann, R. Law and J.A.J. Metz, Cambridge, 2000), pp. 359-387.

\bibitem{keeling}
M.J. Keeling, D.A. Rand and A.J. Morris, Proc. R. Soc. Lond. B {\bf 264}, 1149 (1997).

\bibitem{petermann}
T. Petermann and P. De Los Rios, submitted to J. Theor. Biol. (2003), e-print q-bio.PE/0401028.

\bibitem{diekmann}
O. Diekmann and J.A.P. Heesterbeek, {\it Mathematical Epidemiology of Infectious Diseases} (Chichester, 2000).

\bibitem{molloy1}
M. Molloy and B. Reed, Random Struct. Algorithms {\bf 6}, 161 (1995).

\bibitem{molloy2}
M. Molloy and B. Reed, Combinatorics, Probab. Comput. {\bf 7}, 295 (1998).

\bibitem{maslov}
S. Maslov and K. Sneppen, Science {\bf 296}, 910 (2002).

\bibitem{gross}
J. Gross and J. Yellen, {\it Graph Theory and its Applications} (CRC Press, 1999), p.154.

\end{thebibliography}
\end{document}